\documentclass[a4paper,man,natbib,floatsintext]{apa6}
\usepackage{apacite}
\usepackage{subfig}
\usepackage[english]{babel}
\usepackage[utf8x]{inputenc}
\usepackage{amsmath}
\usepackage{graphicx}
\graphicspath{ {./images/} }
\usepackage[colorinlistoftodos]{todonotes}
\usepackage[left]{lineno}
\usepackage{diagbox}
\usepackage{indentfirst}
\usepackage{adjustbox}
\usepackage{xcolor}
\usepackage{csquotes}
\usepackage{natbib}
\usepackage{tikz}

\usepackage{tablefootnote}
\usepackage{url}

\usetikzlibrary{shapes.geometric, arrows}

\usepackage{caption}
\begin{document}

\begin{titlepage}

\begin{center}

\large\textbf{Toward quantifying trust dynamics: How people adjust their trust after moment-to-moment interaction with automation}


\normalsize

\vspace{35pt}
X. Jessie Yang\\
Industrial and Operations Engineering, University of Michigan\\
\vspace{15pt}
Christopher Schemanske \\
Industrial and Operations Engineering, University of Michigan\\

\vspace{15pt}
Christine Searle \\
Robotics Institute, University of Michigan\\


\end{center}
\begin{flushleft}
\vspace{30pt}
 \textbf{\textit{Accepted to be published in Human Factors 06/29/2021}}

\textbf{Manuscript type:} \textit{Research Article}\\
\textbf{Running head:} \textit{How people adjust their trust in automation}\\
\textbf{Word count:} 4600\\

\textbf{Corresponding author:} X. Jessie Yang, Industrial and Operations Engineering, University of Michigan, Ann Arbor. Email: xijyang@umich.edu\\

\textbf{Acknowledgment:} This material is based upon work supported in part by the National Science Foundation under Grant No. 2045009. 

\vspace{20pt}
\textbf{Pr\'ecis:} We examine how human operators adjust their trust in automation as a result of moment-to-moment interaction with automation. Results indicate that operators' trust adjustments are significantly influenced by decision-making heuristics/biases including the outcome bias and the contrast effect. Additionally, automation failures engender a larger effect on trust adjustment than successes.

\vspace{20pt}
\textbf{Topic:} Automation, Expert Systems

\end{flushleft}

\end{titlepage}
\shorttitle{}

\section{ABSTRACT}

{\textbf{Objective:} We examine how human operators adjust their trust in automation as a result of their moment-to-moment interaction with automation. 

\textbf{Background:} Most existing studies measured trust by administering questionnaires at the end of an experiment. Only a limited number of studies viewed trust as a dynamic variable that can strengthen or decay over time. 

\textbf{Method:} Seventy-five participants took part in an aided memory recognition task. In the task, participants viewed a series of images and later on performed 40 trials of the recognition task to identify a target image when it was presented with a distractor. In each trial, participants performed the initial recognition by themselves, received a recommendation from an automated decision aid, and performed the final recognition. After each trial, participants reported their trust on a visual analog scale. 

\textbf{Results:} Outcome bias and contrast effect significantly influence human operators' trust adjustments. An automation failure leads to a larger trust decrement if the final outcome is undesirable, and a marginally larger trust decrement if the human operator succeeds the task by him-/her-self. An automation success engenders a greater trust increment if the human operator fails the task. Additionally, automation failures have a larger effect on trust adjustment than automation successes.

\textbf{Conclusion:} Human operators adjust their trust in automation as a result of their moment-to-moment interaction with automation. Their trust adjustments are significantly influenced by decision-making heuristics/biases.

\textbf{Application:} Understanding the trust adjustment process enables accurate prediction of the operators' moment-to-moment trust in automation and informs the design of trust-aware adaptive automation.


\textbf{Keywords:} Human-automation interaction, human-autonomy interaction, heuristics and biases, decision aid

\vspace{20pt}

}

\newpage
\section{1. INTRODUCTION}
\noindent Consider the following hypothetical scenarios:
\begin{displayquote}
Scenario A: Assume Mark and Brian were identical from the medical perspective. Both of them spotted blood in their stools. A clinical decision system was used to decide whether they were at risk of colon cancer and if follow-up colonoscopy examinations were necessary. The decision system decided that both patients were at very low risk of developing colon cancer and colonoscopy examinations were unnecessary.
Mark, a hypochondriac fully covered by medical insurance, still requested a follow-up colonoscopy examination, which turned out to reveal cancerous polyps in his colon. A polypectomy afterward removed those polyps and saved his life. In contrast, Brian, constrained by his financial status, did not request a colonoscopy. Shortly after, unfortunately, he was diagnosed with colon cancer. 

\end{displayquote}

\noindent In Scenario A, the clinical decision system made wrong diagnoses for both patients. Therefore, we would expect a decrement of trust toward the system in both cases, but would the levels of trust drop be equal?

\begin{displayquote}


Scenario B: Assume Amy and Marina were identical from the medical perspective. Both Amy and Marina spotted a painful lesion on their skins. Their symptoms were analyzed by a medical doctor and a clinical decision system. Based on the medical doctor’s diagnoses, only Amy has developed skin cancer. However, the clinical decision system concluded that both patients had skin cancer. A tandem expert review confirmed that the decisions by the clinical system were correct. 

\end{displayquote}

\noindent In Scenario B, the clinical decision system made correct diagnoses for both patients. Therefore we expect an increment of trust toward the system in both cases, but would the levels of trust increment be equal? 

\newpage
The present study seeks to answer these questions. We begin by reviewing existing literature on trust adjustment in human-automation interaction. Next, we discuss the two types of decision-making heuristics/biases, namely outcome bias and contrast effect. We hypothesize that outcome bias and contrast effect would affect human operators' trust adjustments when interacting with automation. Although there are other types of decision-making heuristics/biases, these two types are particularly relevant to trust adjustment and are our focus in the present study. 

\subsection{1.1 Trust as a Dynamic Variable}

\begin{figure}[H]
  \begin{center}
  \vspace{0pt}
    \includegraphics[width=0.6\textwidth]{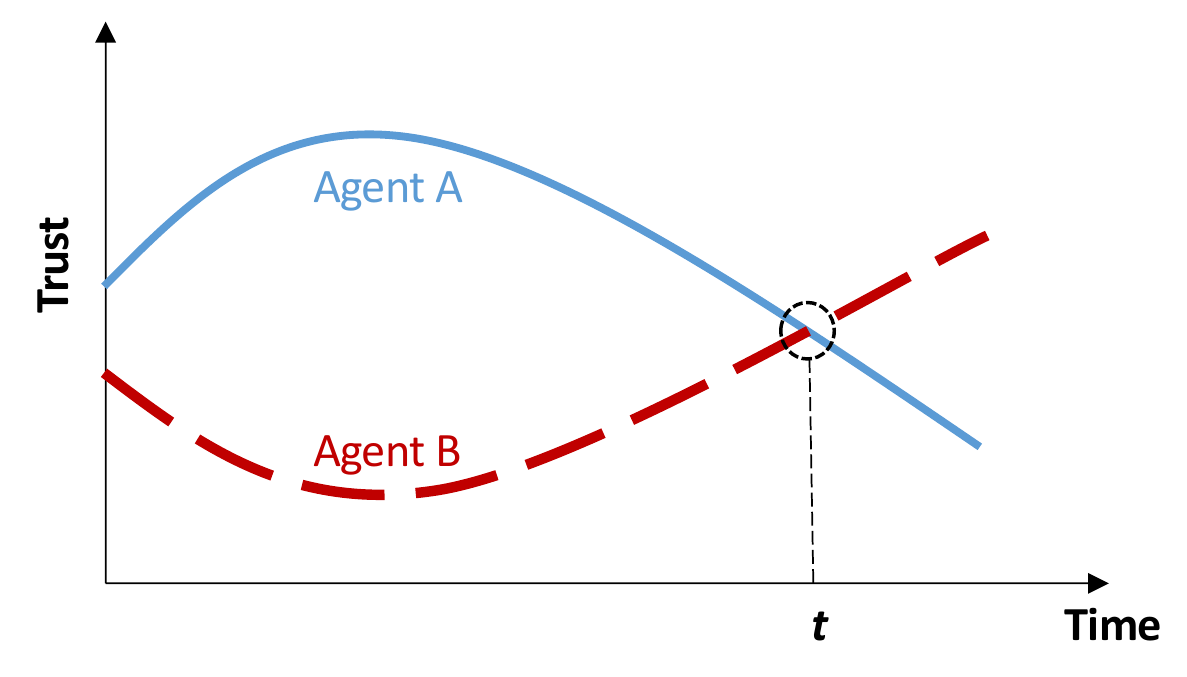}
  \end{center}
  \vspace{-20pt}
  \caption{The static \textit{snapshot} view of trust versus the \textit{dynamic} view of trust. If taking a snapshot at time $t$, both agents have the same trust level, but their trust dynamics differ.}
  \label{fig:snapshot}
\end{figure}

Trust in automation, or more recently, trust in autonomy, has attracted substantial research attention in the past three decades. The majority of prior literature adopted a \textit{snapshot} view of trust and typically evaluated trust through questionnaires administered at the end of an experiment. More than two dozen factors have been identified to influence one's (snapshot) trust in automation, including individual factors such as culture and age \citep{Rau2009, McBride:2011ix}, system factors such as reliability and level of automation \citep{Wickens:2007hm, Wickens:2009False,Parasuraman:2000, Du2019}, and environmental factors such as multi-tasking requirement \citep{Zhang2017:HFES}. 
This snapshot view, however, does not fully acknowledge that trust is a dynamic variable that can change over time (Figure \ref{fig:snapshot}). With few exceptions, we have little understanding of how trust strengthens or decays due to moment-to-moment interactions with automation \citep{Yang:2017:EEU:2909824.3020230, DeVisser_IJSR, Guo2020_IJSR, Yang2021}. 

The limited amount of research on trust adjustment reveals two major findings. First, human operators' trust adjustments are significantly influenced by automation performance: trust increases after automation successes and decreases after automation failures \citep{Lee:1992it,Moray:2000,Yang:2017:EEU:2909824.3020230, Yang:2016}. Second, automation failures engender a much stronger influence on trust adjustment than automation successes -- Trust is difficult to build but can be lost quickly \citep{manzey2012human, Lee:1992it, lee1994trust}. 

In their seminal work, \citet{Lee:1992it} employed a simulated pasteurization task, in which participants controlled two pumps and one heater, each of which could be set to automatic or manual control. Participants completed 10 training trials and 50 experimental trials over three days. Two pump faults were introduced on trials 26 and 40, at which point the pump failed to respond accurately. After each trial, participants rated their level of trust on a 10-point Likert scale. Trust plotted over the 60 trials showed a mild trust increment after interacting with a reliable pump but a large trust decrement after trials 26 and 40. Based on the results, \citet{Lee:1992it} developed an autoregressive time-series model of trust. Trust at time $t$ was modeled as a function of trust at time $t-1$, the pasteurization output, and the occurrence of pump faults.

Along the same line, \citet{manzey2012human} used the AutoCAMS multi-tasking platform \citep{Hockey1998} 
to track the human operators' trust over 20 positive interactions (automation successes) and 1 or 2 negative interactions (automation failures). Participants rated their subjective trust 5 times throughout the experiment. Results of the experiment reveal a sharp decline of trust after the automation failures. \citet{manzey2012human} concluded that human operators adjust their trust in automation based on positive and negative feedback loops. The positive loop is triggered by the experience of automation success, and the negative loop by the experience of automation failures. 

Recently, \citet{Yang2021} summarized the two major findings as two properties of trust dynamics, namely \textit{continuity} and \textit{negativity bias}, and identified the third property, \textit{stabilization} (i.e., An average person's trust will stabilize over repeated interactions with the same automation.) Based on the three properties, a computational model for predicting the moment-to-moment trust was developed \citep{Guo2020_IJSR, Yang2021, Guo2021_RAL}. The computational model proposes that trust at any time, $t$, follows a Beta distribution. The model outperforms existing models in prediction accuracy and guarantees good model e generalizability and explainability.

\subsection{1.2 Decision Making Heuristics/Biases and Trust Adjustment}


When making decisions, people rarely use the normative approach. Instead, their decision-making is often subject to various types of heuristics/biases. Below we discuss three types of decision-making heuristics/biases, namely, hindsight bias, outcome bias, and contrast effect. 

Hindsight bias, as described by \citet{fischhoff1975hindsight}, is the tendency to adjust the estimates of various likelihoods of possible event outcomes in uncertain situations after the event has occurred and the outcome is known. Hindsight bias is also known as the creeping determinism \citep{fischhoff1975hindsight} and knew-it-all-along effect \citep{wood1978knew}. It has been documented in auditory processing, labor disputes, medical diagnoses, consumer satisfaction, personnel management, sporting events, political strategy, legal proceedings, and nuclear accident analysis \citep{bernstein2012auditory, roese2012hindsight, hawkins1990hindsight}. The typical experimental procedure presents participants with a situation that may lead to several possible outcomes. Participants estimate the likelihood of each outcome, learn the actual outcome, and estimate the likelihood of each outcome again \citep{guilbault2004meta}. Hindsight bias occurs when the participants rate the actual outcome as more likely in the second estimate and the other possible outcomes as less likely. 

Rather than observing a change in the estimated likelihood of each outcome, outcome bias observes a change in the perceived quality of the decision made \citep{baron1988outcome, henriksen2003hindsight}. Outcome bias has been observed in laboratory tasks evaluating medical decisions \citep{baron1988outcome}, gambling \citep{baron1988outcome, brownback2019understanding, sezer2016overcoming}, and business decisions \citep{gino2009see}, and in real-world evaluations of soccer player performance \citep{kausel2019outcome}. Following a similar experimental paradigm, outcome bias studies ask participants to rate the quality of a decision that influences but do not entirely determine outcomes. After that, participants learns the actual outcome and rate the decision quality again. The probabilities of each outcome are held constant and in some cases even made explicit to participants \citep{baron1988outcome, brownback2019understanding}, suggesting that any differences between the pre- and post-outcome decision quality judgments can be attributed to the only new information provided -- the outcome \citep{baron1988outcome}. Outcome bias occurs when the same decision is evaluated to be of lower quality when it happens to produce bad rather than good outcome. 



Another decision making heuristics/biases people may use in trust adjustment is the contrast effect. The contrast effect occurs when people's judgments are unintentionally affected by previous or simultaneous stimuli. It has been observed in judgments of shape perception \citep{suzuki1998shape}, facial recognition \citep{hsu2016relative}, job candidate interviews \citep{wexley1972importance}, physical attractiveness \citep{thornton1997physique, kenrick1980contrast}, and consumer reports \citep{lynch1991contrast}. In their study of sequential effects on perceptions of job candidates, \citet{wexley1972importance} found that up to 12\% of variance in ratings of candidates could be attributed to a contrast effect from the two candidates preceding the target candidate.

\subsection{1.3 The Present Study}

The primary objectives of the present study are two-folded. First, we aim to provide further evidence on the effects of automation successes and failures on a person's trust adjustment. Prior studies employing a small number of automation failures showed that a person's moment-to-moment trust increases after automation successes and decreases after automation failures \citep{Lee:1992it,Moray:2000,Yang:2017:EEU:2909824.3020230, Yang:2016}, and automation failures have a larger influence on trust adjustment \citep{manzey2012human, Lee:1992it, lee1994trust}. We will replicate and further examine the two empirical findings with more frequent occurrences of automation failures. 


Second, we aim to investigate the effects of outcome bias and contrast effect on trust adjustment. The two types of biases/heuristics have been observed in various judgment and decision-making tasks \citep{baron1988outcome, henriksen2003hindsight, suzuki1998shape, hsu2016relative}. In the context of trust adjustment in human-automation interaction, we postulate that people’s moment-to-moment trust adjustment will be affected by the final outcome of a task and will be influenced by whether a task could be completed successfully by a human operator him-/her-self manually. To our knowledge this is the first experiment to examine how outcome bias and contrast effect influence people’s moment-to-moment trust adjustment. In particular, we test the following hypotheses:


\noindent\textbf{\textit{H1a}}: Trust in automation will increase as a result of automation successes and decrease as a result of automation failures. 

\noindent\textbf{\textit{H1b}}: The magnitude of trust decrements due to automation failures will be greater than that of trust increments due to automation successes.

\noindent\textbf{\textit{H2a}}: An automation success will lead to a larger increment of trust if the final outcome of a task is desirable. 

\noindent\textbf{\textit{H2b}}: An automation failure will lead to a larger decrement of trust, if the final outcome of a task is undesirable.

\noindent\textbf{\textit{H3a}}: An automation success will lead to a greater increment of trust if a human operator fails the task on his or her own.

\noindent\textbf{\textit{H3b}}: An automation failure will lead to a greater decrement of trust if a human operator succeeds the task on his or her own.





Along with the primary objectives, we were interested in exploring any potential impact of the overall automation reliability on trust adjustment. Previous research has revealed consistently a positive relationship between automation reliability and the (snapshot) trust in automation measured at the end of an experiment \citep{ DeVisser:2011dg, Wickens:2007hm}. 
However, little research, if any, has examined the effects of automation reliability on moment-to-moment trust adjustment.  In the present study, automation reliability was set to be above 70\% (i.e., 70\%, 80\% and 90\%) based on a meta-analysis showing that 70\% is the threshold above which using automation improves task performance and below which such use may harm performance \citep{Wickens:2007hm}.

\section{2. METHOD}
This research complied with the American Psychological Association code of ethics and was approved by the Institutional Review Board at the University of Michigan.

\subsection{2.1 Participants}
A total of 75 adults (Age: Mean = 23.4 years, \textit{SD} = 4.1 years) participated. The number of participants was determined using a power analysis for the 2 (pattern) $\times$ 3 (reliability level) mixed design $F$ test. The power analysis was completed by assuming a large effect, a $\alpha$ of .05, and a statistical power of .80. The required sample size is 51. All participants had normal or corrected-to-normal vision.  Participants were paid a base rate of \$10 dollars plus a bonus ranging from \$1 to \$5 depending on their performance.   

Of the 75 participants, 60 performed the experiment in a face-to-face setting and the remaining 15 in a remote control setting where they controlled the experimental PC remotely (due to the COVID-19 pandemic). We notice no systematic differences in participants' behavior or performance between the two groups except that the participants in the remote control setting took longer to complete the experiment, probably due to internet lags when controlling the experimental PC remotely.

\subsection{2.2 Apparatus and Stimuli}
\begin{figure}[H]
\centering
\includegraphics[width= 0.8\textwidth]{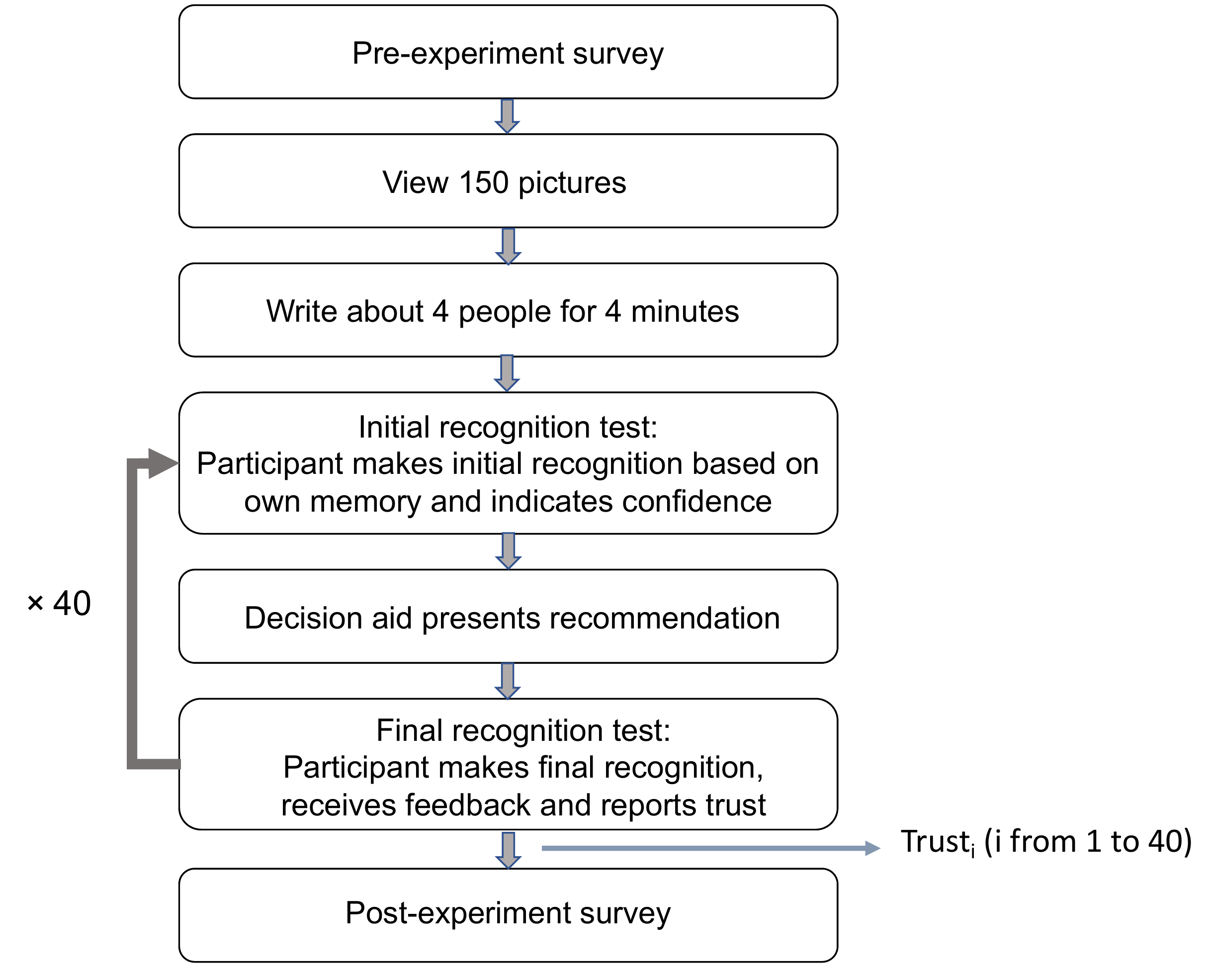}
\caption{Flow chart of the aided memory recognition task}
\label{fig:flowchart}
\end{figure}

The study employs a simulated memory recognition task adapted from \citet{tulving1981}. Figure \ref{fig:flowchart} shows the flowchart of the experimental task.

Before the experiment, participants fill in a demographic survey (i.e., age and gender) and a trust propensity survey gauging their propensity to trust automation. Trust propensity has been shown to influence a person's (snapshot) trust in automation after interacting with an automated system  \citep{Merritt:2008ds}. Please refer to Appendix A for the items used in the survey.

\begin{figure}[H]
\centering
\includegraphics[width= 0.8 \textwidth]{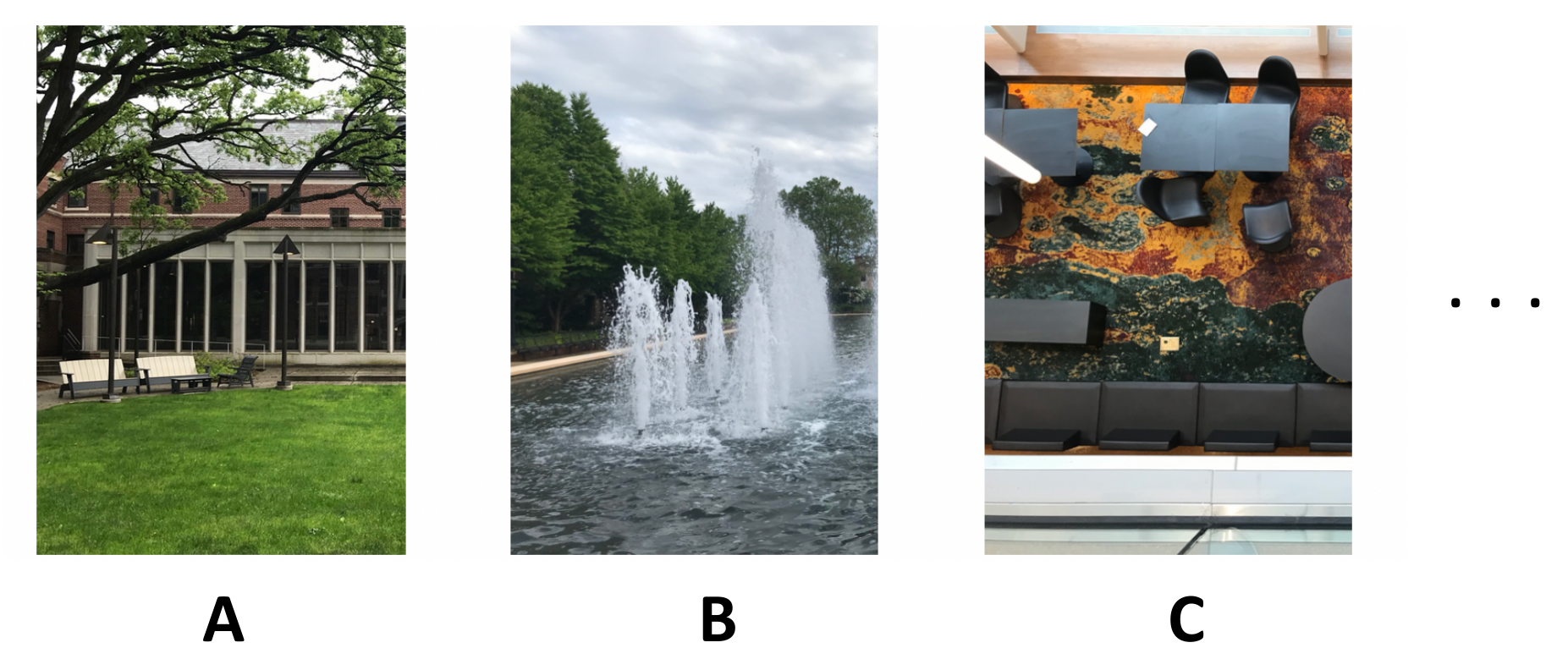}
\caption{Illustration of target pictures A, B, C, ...}
\label{fig:ABC}
\vspace{-5mm}
\end{figure}

In the experiment, each participant first views a block of 150 pictures (i.e., A, B, C, ...), each for two seconds, as shown in Figure \ref{fig:ABC}. 
Sixty of the 150 images are targets for later recognition, and 90 are buffer images to increase cognitive load.  
Next, participants perform an interpolated memory task to write down as much information as they could about four famous people in four minutes. The interpolated task is used in the study of \citet{tulving1981} to bring the hit rate into the middle performance range. The number of famous people and the duration of the interpolated task are determined in a preliminary study \citep{Yang:2016}.

\begin{figure}[H]
\centering
\includegraphics[width= 0.6 \textwidth]{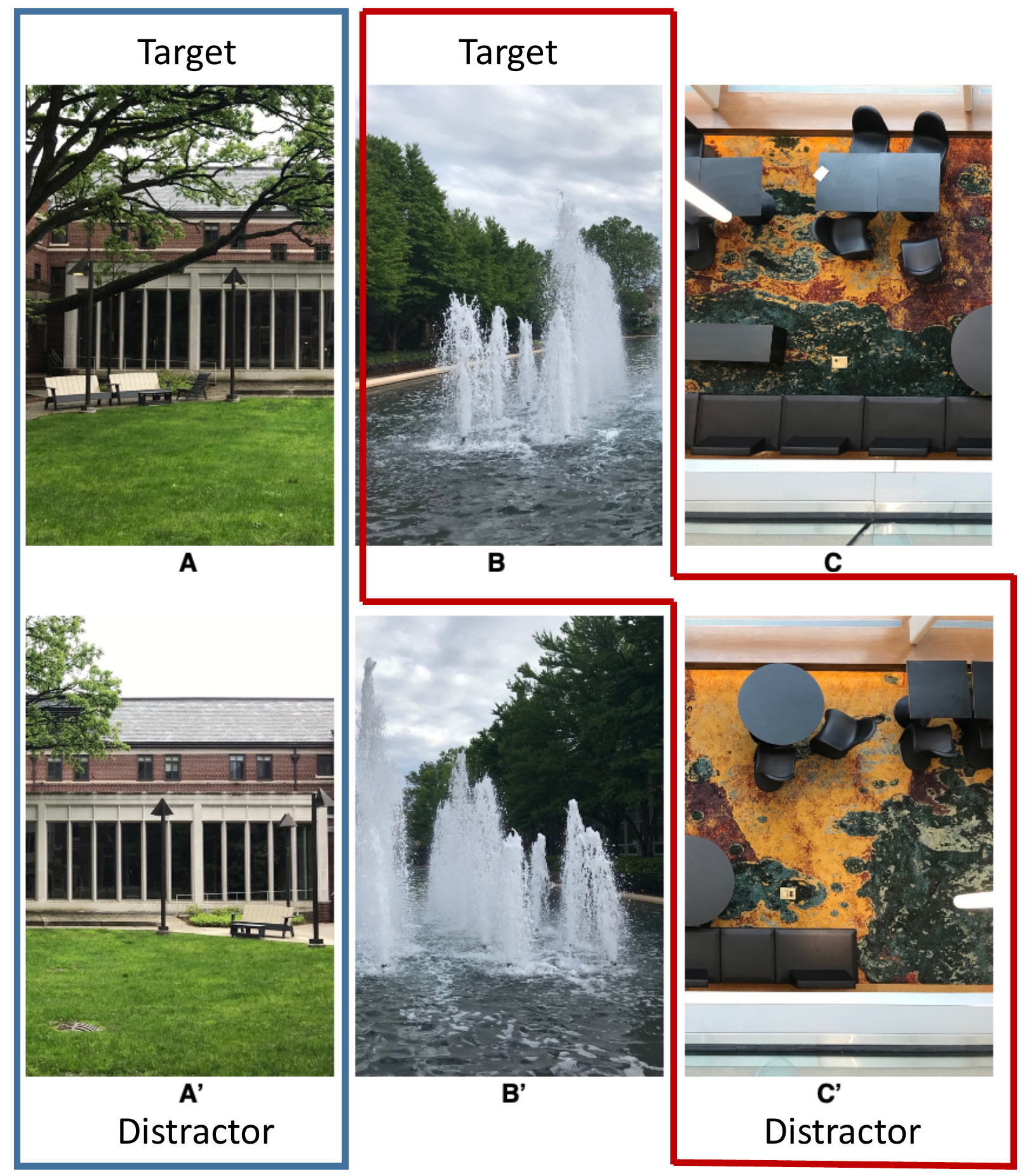}
\caption{Illustration of target-distractor pairs}
\label{fig:create40}
\vspace{-5mm}
\end{figure}

Participants are then given the recognition test that consists of 40 trials of a two alternative forced choice image recognition task (2AFC), in which participants identify a target image when it is presented with a distractor. The 40 trials of target-distractor pairs are created from the 60 target pictures (Figure \ref{fig:create40}). In each trial, a target picture could be presented with a distractor that resembles itself (e.g, Distractor A' resembles target A) or a target picture could be presented with a distractor that resembles another target picture (e.g., Distractor C' resembles target C). Therefore, 60 target pictures only generate 40 target-distractor pairs.

\begin{figure}[H]
\centering
\includegraphics[width= 0.8 \textwidth]{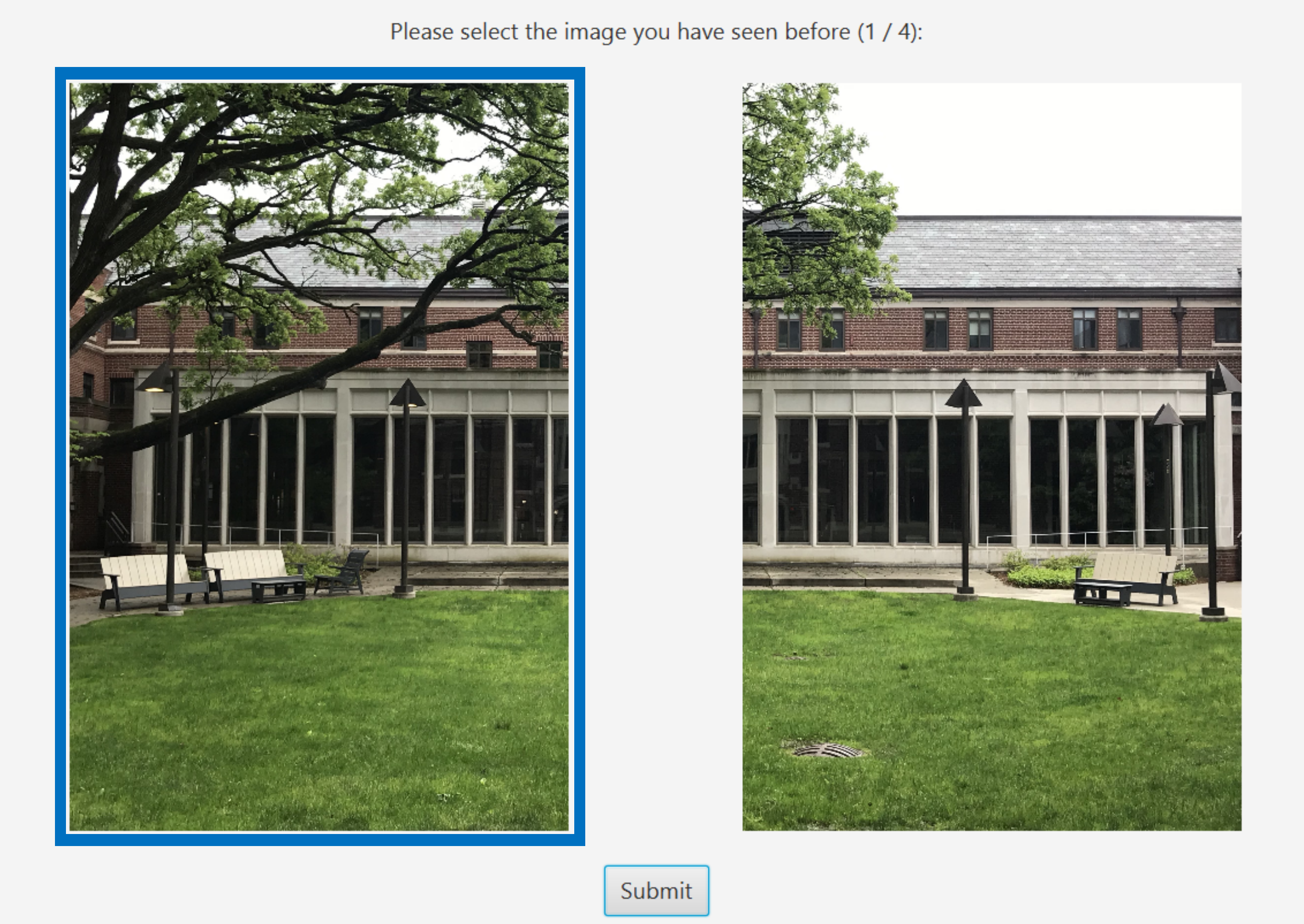}
\caption{Illustration of the two-alternative forced-choice (2AFC) test}
\label{fig:2AFC}
\vspace{-5mm}
\end{figure}

During each recognition trial, participants first selects the image they recall seeing previously entirely based on their memory by clicking on it with the cursor (Figure \ref{fig:2AFC}). The selected image will be highlighted in a frame. Participants then rate their confidence using a visual analog scale, with the leftmost point labeled ``I'm completely guessing.'' and the rightmost point ``I'm completely certain.''

\begin{figure}[H]
\centering
\includegraphics[width= 0.4\textwidth]{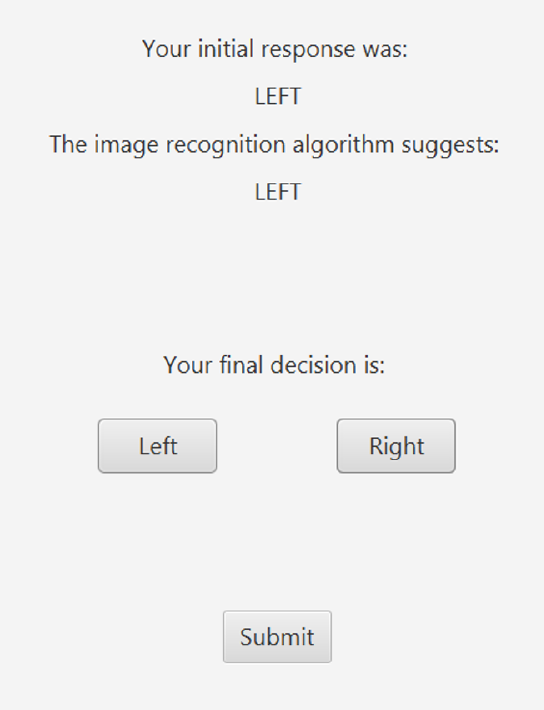}
\caption{The automated aid's recommendation interface}
\label{fig:recc}
\vspace{-5mm}
\end{figure}

After that, the participants are reminded of their initial choice, and presented with the recommendation from an automated decision aid (i.e., displaying ``The image recognition algorithm suggests LEFT/RIGHT'' on screen). The participants are then asked to make their final recognition selection (Figure \ref{fig:recc}). 
Once the participants make their final recognition selection, they receive feedback on the correctness of their final choice (i.e., ``Your final answer is CORRECT/WRONG'' on screen). 

\begin{figure}[H]
\centering
\includegraphics[width= 0.8 \textwidth]{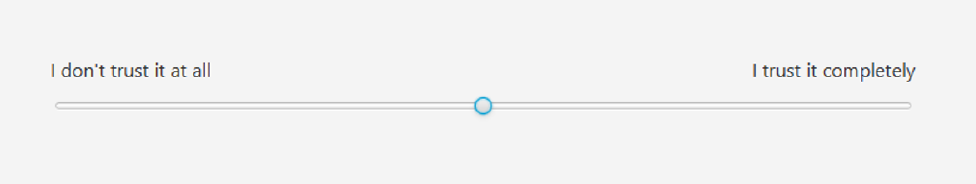}
\caption{Illustration of the visual analog scale for measuring $trust_i$ after each trial}
\label{fig:slider}
\vspace{-5mm}
\end{figure}

After that, participants rate their trust towards the automated decision aid using a visual analog scale \citep{manzey2012human} (Figure \ref{fig:slider}), with the leftmost point labelled ``I don't trust it at all.'' and the rightmost point ``I trust it completely.''  All visual analog scales are then converted to 0-100 scales for data analysis. 

After participants complete the experiment, they fill in two post-experiment trust survey adapted from \citet{Jian:2000es} and \citet{MORAY:1996it}. Please refer to Appendixes B and C for the items used in the two surveys.

\subsection{2.3 Experimental Design}
The experiment employs a mixed design with two independent variables. The between-subjects variable is automation reliability, varied in three levels: 70\%, 80\%, and 90\%. The within-subjects variable is performance pattern. In the experiment, participants perform an initial recognition, received a recommendation from the automated decision aid, and performed a final recognition. All the three steps could be either correct or wrong, resulting in 8 performance patterns (Table \ref{tab:8 patterns with data}). For example, Pattern 3 indicates that the participants make a wrong recognition initially (i.e., 0 in binary format), but after receiving a correct recommendation from the automated aid (i.e., 1 in binary format), change the answer, and the final recognition is correct (i.e., 1 in binary format). Note that each participant does not necessarily exhibit each of the eight patterns because the performance patterns are a result of their recognition performance.




\subsection{2.4 Measures}
\indent \textbf{\textit{Trust propensity.}} At the beginning of the experiment, participants complete a survey gauging their propensity to trust automation, adapted from \citet{merritt2013}.

\indent \textbf{\textit{Trust adjustment.}} 
After each 2AFC trial $i$,  participants report their $trust(i)$ in the decision aid. We calculate a trust adjustment as:

\begin{center}

$Trust \ adjustment (i) = Trust (i) -  Trust (i-1)$, where $i = 2, 3,...,40$
\end{center}

Since the moment-to-moment trust is reported after each trial, only 39 trust adjustments are obtained from each participant.

\indent \textbf{\textit{Post-experiment Trust Survey.}} After the experiment, participants complete a 12-item trust survey  \citep{Jian:2000es} and an 8-item trust survey \citep{MORAY:1996it}.





\subsection{2.5 Experimental Procedure}
Prior to the experiment, participants provided informed consent and completed a demographic survey and the trust propensity survey. They were oriented to steps of the experiment and walked through each of the screens they would see during the experiment. After that, participants completed a practice session, wherein they viewed 12 images, performed the interpolated four well-known-people memory task, and performed four 2AFC trials.  Participants then proceeded to the experimental trials, viewed 150 new images, performed the interpolated task, and completed 40 2AFC trials with their assigned automation reliability level. The images and the well-known people used in the practice session were different from the ones in the actual experiment. At the end of the experiment, participants reported their post-experiment trust toward the automated aid using two scales \citep{Jian:2000es, MORAY:1996it}. Participants were told that the automated decision aid was imperfect, but they were not informed of the exact reliability level. 
The experiment spanned roughly half an hour, and the average time for the 40 2AFC test trials was 8 minutes and 54 seconds (SD = 1 minute and 58 seconds)  for the in-person group, and 13 minutes and 38 seconds (SD = 5 minutes and 35 seconds) for the remote group.


\section{3. RESULTS}

After the experiment was completed, the number of occurrences for each performance pattern was calculated. Table \ref{tab:8 patterns with data} summarizes the number of occurrences for each pattern and the corresponding trust adjustments. As the occurrence of each pattern can only be determined posteriorly, participants might not necessarily display each performance pattern. 
Due to the extremely low number of occurrences for patterns 1 and 6, these two patterns are discarded from data analysis.  As a result, a full factorial analysis
is inappropriate. Instead, we conduct a series of planned comparisons. For the planned comparisons, we first conduct Analysis of Covariance (ANCOVAs) with both automation reliability and performance pattern as independent variables, and trust propensity as the covariate.  Results show that neither automation reliability (i.e., 70\%, 80\%, 90\%) or trust propensity had significant effects nor interaction effects in all the planned comparisons. Therefore, trust propensity (as the covariate) is removed from the analysis and the data are collapsed across the levels of automation reliability. The following analysis is conducted on the collapsed dataset using one sample and paired samples t-tests. Data points outside of three standard deviations of the mean are considered outliers and excluded from the data analysis.

\begin{table}[H]
\caption{8 ($2 \times 2 \times 2$) possible performance patterns based on the combinations of human operator's initial recognition, the recommendation provided by the automated decision aid, and the operator's final recognition}
\begin{adjustbox}{width=1\textwidth}
\begin{tabular}{cccccc}
\hline
\textbf{\begin{tabular}[c]{@{}c@{}}Performance Pattern \\ Decimal: Binary\end{tabular}} & \textbf{Initial Recognition} & \textbf{Recommendation} & \textbf{Final Recognition} & \textbf{\# of Participants} & \textbf{\begin{tabular}[c]{@{}c@{}}Trust Adjustment\\ Mean (SD)\end{tabular}} \\ \hline
0: 000 & Wrong (0) & Wrong (0) & Wrong (0) & 68 & -4.2 (4.2) \\
\textsuperscript{$\ast$} 1: 001 & Wrong (0) & Wrong (0) & Correct (1) & 1 & NA \\
2: 010 & Wrong (0) & Correct (1) & Wrong (0) & 72 & 1.6 (1.7) \\
3: 011 & Wrong (0) & Correct (1) & Correct (1) & 71 & 1.9 (1.3) \\
4: 100 & Correct (1) & Wrong (0) & Wrong (0) & 63 & -5.0 (4.6) \\
5: 101 & Correct (1) & Wrong (0) & Correct (1) & 63 & -3.7 (4.4) \\
\textsuperscript{$\ast$} 6 : 110 & Correct (1) & Correct (1) & Wrong (0) & 5 & NA \\
7: 111 & Correct (1) & Correct (1) & Correct (1) & 73 & 1.2 (1.0) \\ \hline
\end{tabular}
\end{adjustbox}
\textit{Note}: Asterisks denote the exclusion of a pattern due to its low number of occurrences.
\label{tab:8 patterns with data}
\end{table}

\begin{figure}[H]
\centering\subfloat{\includegraphics[width=0.6\textwidth]{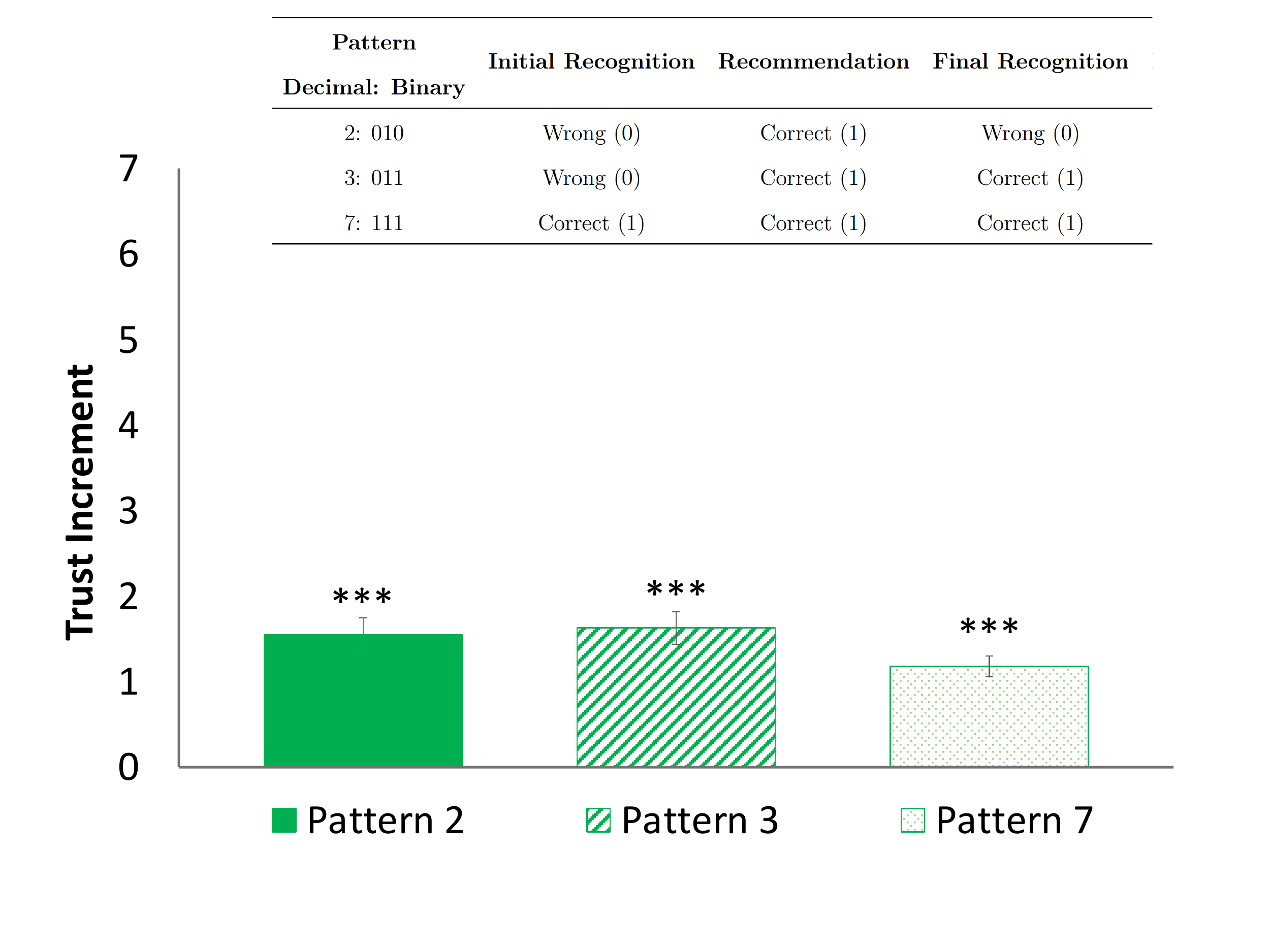}}
\caption{Mean and Standard Error(SE) values of \textbf{\textit{trust increment}} in patterns 2, 3 and 7
(\textsuperscript{$\ast\ast\ast$}$p$  < .001).}
\label{fig: pattern237}
\end{figure}

\begin{figure}[H]
\centering\subfloat{\includegraphics[width=0.6\textwidth]{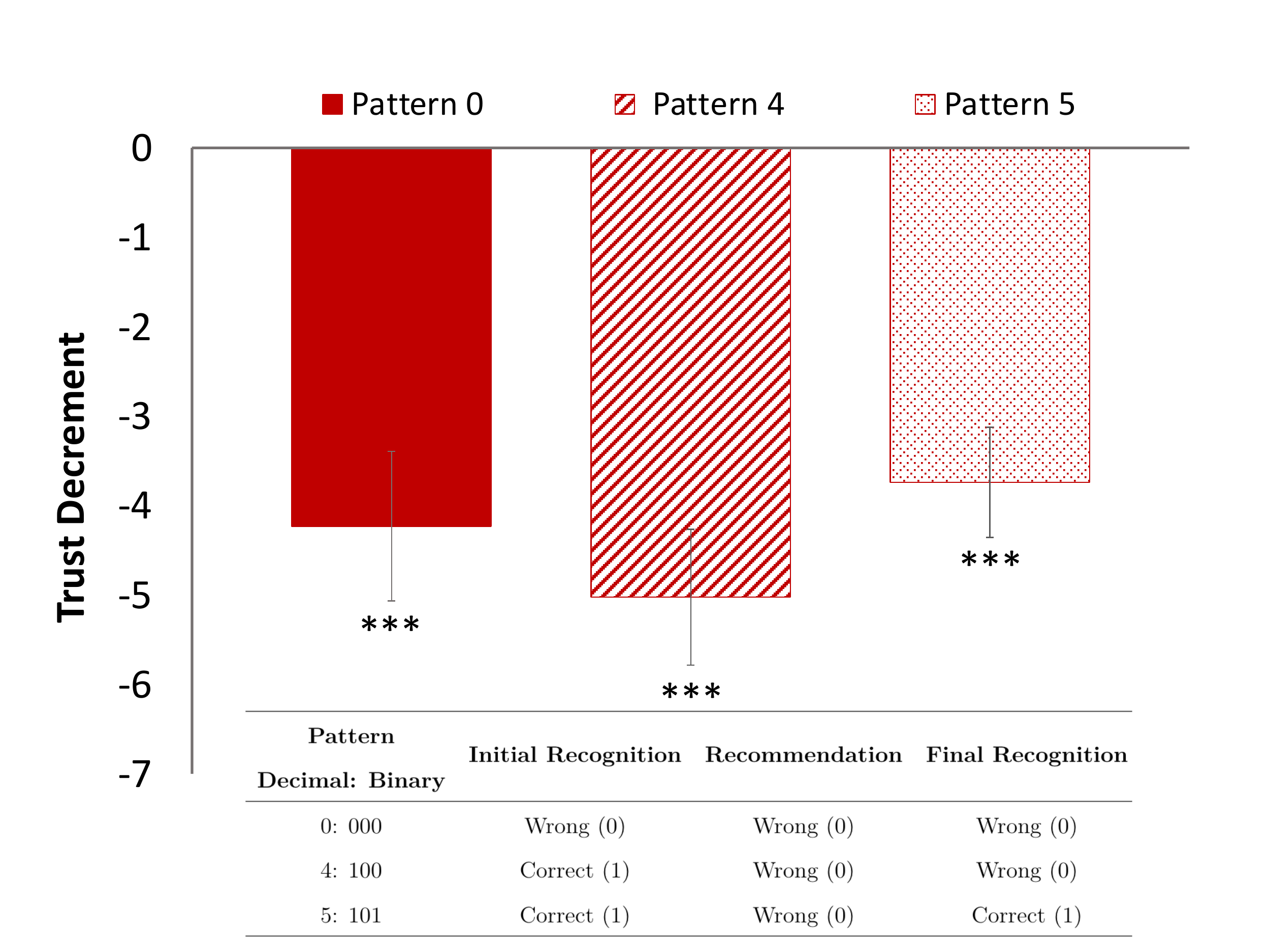}}
\caption{Mean and Standard Error (SE) values of \textbf{\textit{trust decrement}} in patterns 0, 4 and 5  
(\textsuperscript{$\ast\ast\ast$}$p$  < .001).}
\label{fig: pattern045}
\end{figure}

Recalling that \textbf{\textit{H1a}} hypothesizes that an automation success would result in trust increment and an automation failure trust decrement, we compare the magnitude and direction of trust adjustment of patterns 2, 3, and 7 (correct recommendations) against zero, and of patterns 0, 4, and 5 (wrong recommendation) against zero. One sample t-tests with Bonferroni adjustments ($\alpha = 0.017$) show that correct recommendations increase trust (Pattern 2: $t(1, 71) = 7.81, p < 0.001$, Cohen's $d = 0.92$; Pattern 3: $t(1,70) = 12.6, p < 0.001$, Cohen's $d = 1.48$; Pattern 7: $t(1,72) = 9.94, p < 0.001$, Cohen's $d = 1.17$) (Figure \ref{fig: pattern237})  and wrong recommendations decrease trust (Pattern 0: $t(1,67) = -8.29, p < 0.001$, Cohen's $d = 1.00$; Pattern 4: $t(1,62) = -8.72, p < 0.001$, Cohen's $d = 1.10$; Pattern 5: $t(1, 62) = -6.74, p < 0.001$, Cohen's $d = 0.85$) (Figure \ref{fig: pattern045}).


\begin{figure}[H]
\centering\subfloat []{\includegraphics[width=0.47\textwidth]{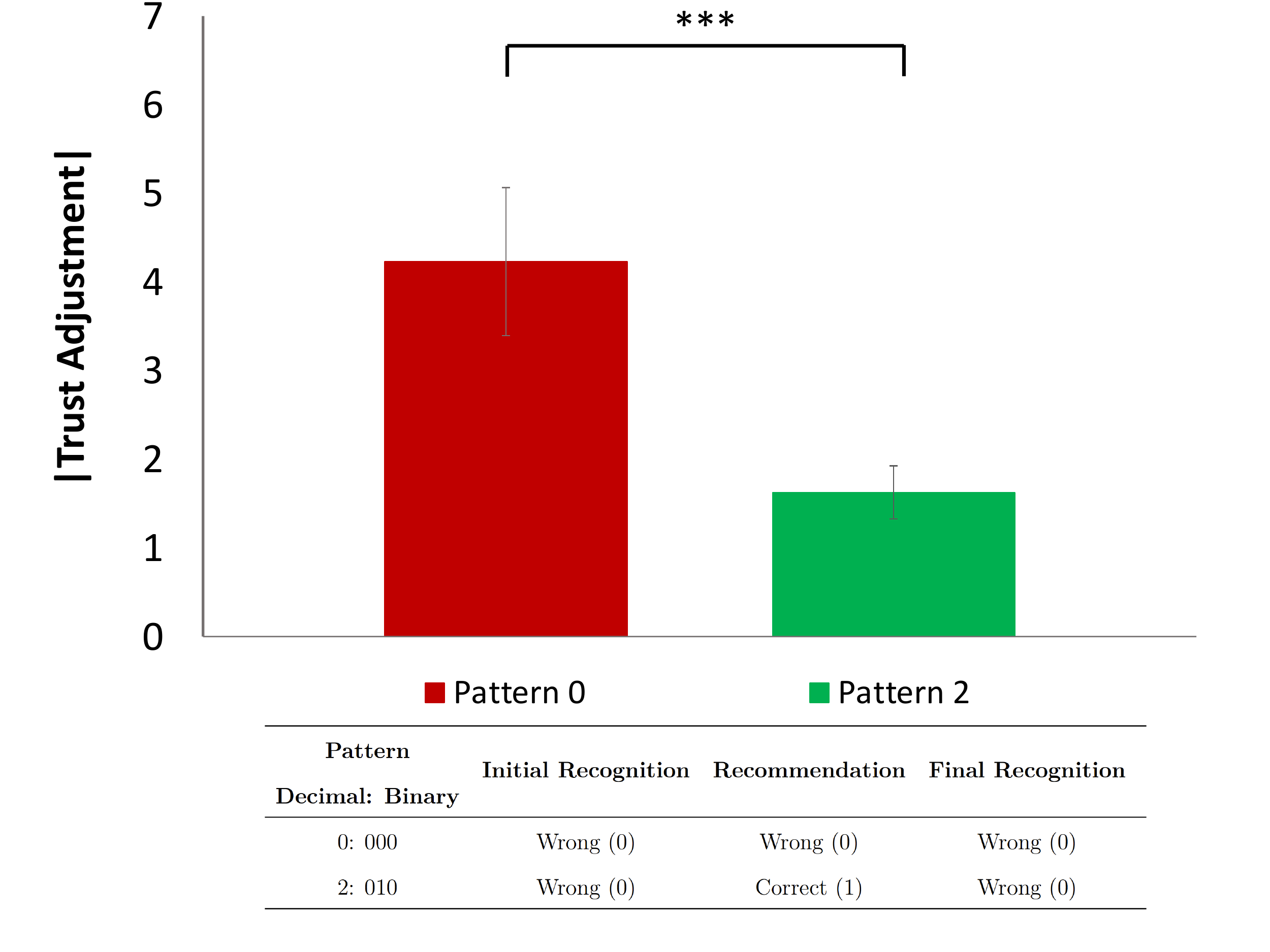}}
\hspace{10pt}
\subfloat []{\includegraphics[width=0.47\textwidth]{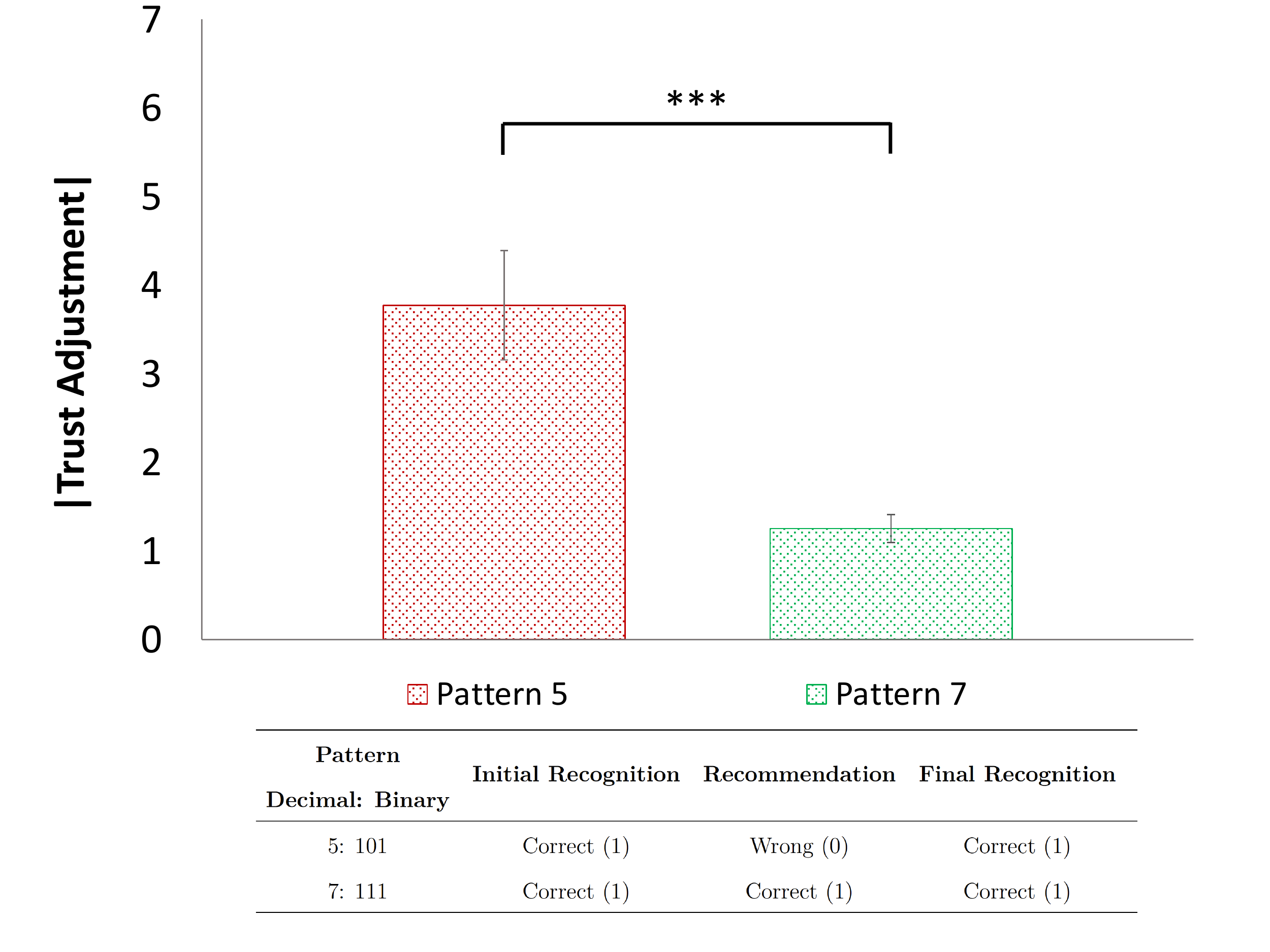}}
\vspace{-5mm}
  \caption{Comparing the \textit{\textbf{magnitude}} of trust adjustment between (a) patterns 0 
  and 2 
  and between (b) patterns 5 
  and 7
(\textsuperscript{$\ast\ast\ast$}$p$  < .001).}
  
  \label{fig: magnitude_comparison}
\end{figure}

\textbf{\textit{H1b}} hypothesizes that the magnitude of trust decrements would be larger than that of trust increments. We conduct two paired samples t-tests with Bonferroni adjustments ($\alpha = 0.025$): the \textit{magnitude} of pattern 0 (wrong initial recognition-wrong recommendation-wrong final recognition) versus the magnitude of pattern 2 (wrong initial recognition-correct recommendation-wrong final recognition), and the magnitude of pattern 5 (correct initial recognition-wrong recommendation-correct final recognition) versus the magnitude of pattern 7 (correct initial recognition-correct recommendation-correct final recognition). 
The only difference between each pair is the correctness of the automation’s recommendation. Results reveal higher magnitude of trust decrements for wrong recommendations in pattern 0 compared to correct recommendations in pattern 2 ($t(1,65) = 5.34, p < 0.001$, Cohen's $d = 0.66$), and for wrong recommendations in pattern 5 compared to correct recommendations in pattern 7 ($t(1,60) = 4.81, p < 0.001$, Cohen's $d = 0.62$) (Figure \ref{fig: magnitude_comparison})


\begin{figure}[H]
\centering\subfloat[]{\includegraphics[width=0.47\textwidth]{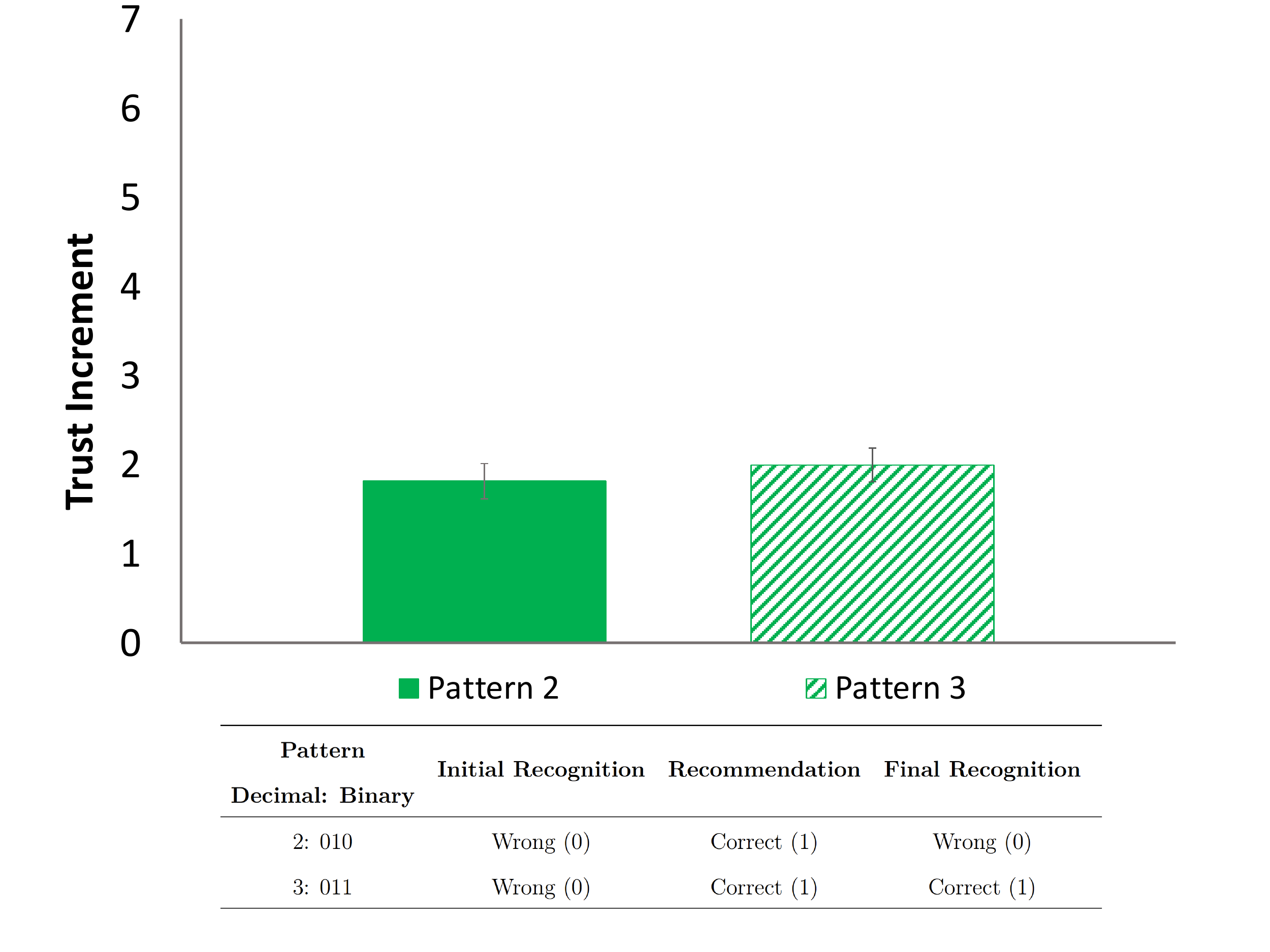}}
\hspace{10pt}
\centering\subfloat[]{\includegraphics[width=0.47\textwidth]{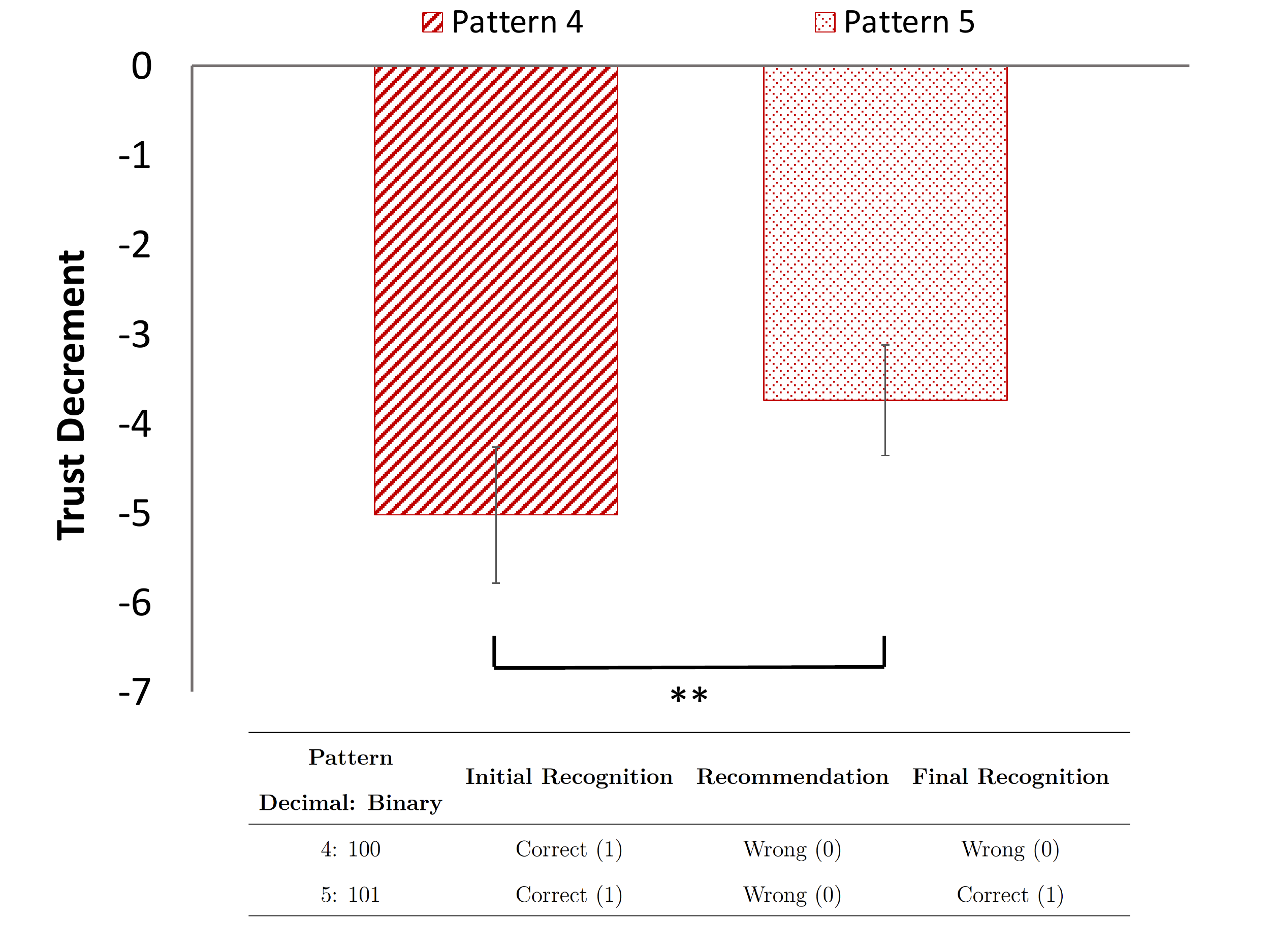}}

\caption{(a) Comparing \textbf{\textit{trust increment}} between patterns 2 
and 3. 
(b) Comparing \textbf{\textit{trust decrement}} between patterns 4 
and 5
(\textsuperscript{$\ast\ast$}$p$  < .01).}
\label{fig: outcome_bias}

\end{figure}

\textbf{\textit{H2a}} hypothesizes that an automation success would lead to a larger trust increment if the final outcome is good. We compare pattern 2 (wrong initial recognition - correct recommendation - wrong final recognition) versus pattern 3 (wrong initial recognition - correct recommendation - correct final recognition), where the only difference is the final recognition performance.  The t-test shows that the difference is not significant ($t(1,67) = -1.08, p = .29$) (Figure \ref{fig: outcome_bias}(a)).




\textbf{\textit{H2b}}  hypothesizes that an automation failure would lead to a larger trust decrement if the final outcome is undesirable. The comparison between pattern 4 (correct initial recognition - wrong recommendation - wrong final recognition) and pattern 5 (correct initial recognition - wrong recommendation - correct final recognition) reveals a significantly larger decrement for pattern 4 (wrong final recognition) ($t(1,52) = -2.63, p = .01$, Cohen's $d = 0.36$) (Figure \ref{fig: outcome_bias}(b)). 

\begin{figure}[H]
\centering\subfloat[]{\includegraphics[width=0.47\textwidth]{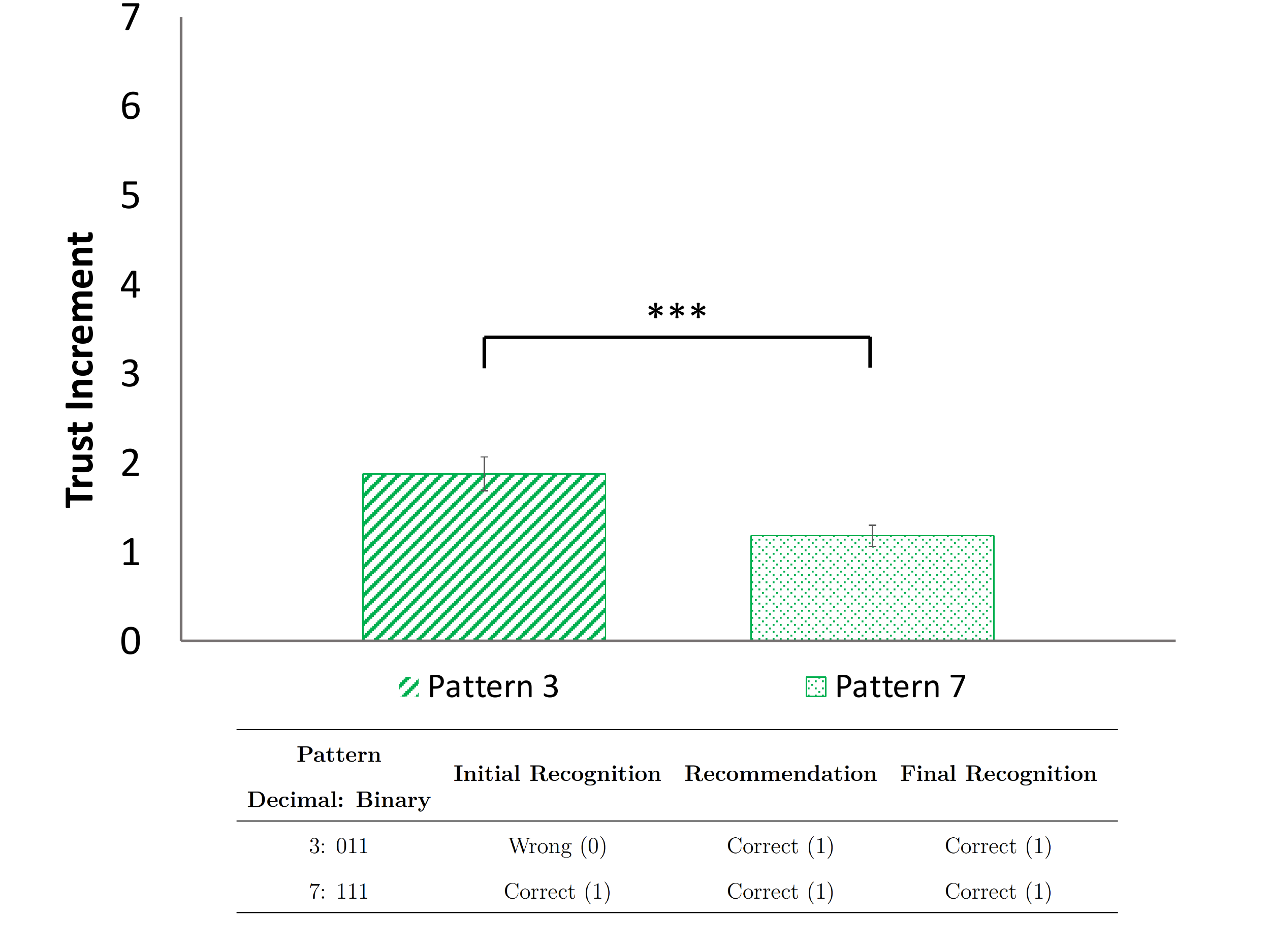}}
\hspace{10pt}
\centering\subfloat[]{\includegraphics[width=0.47\textwidth]{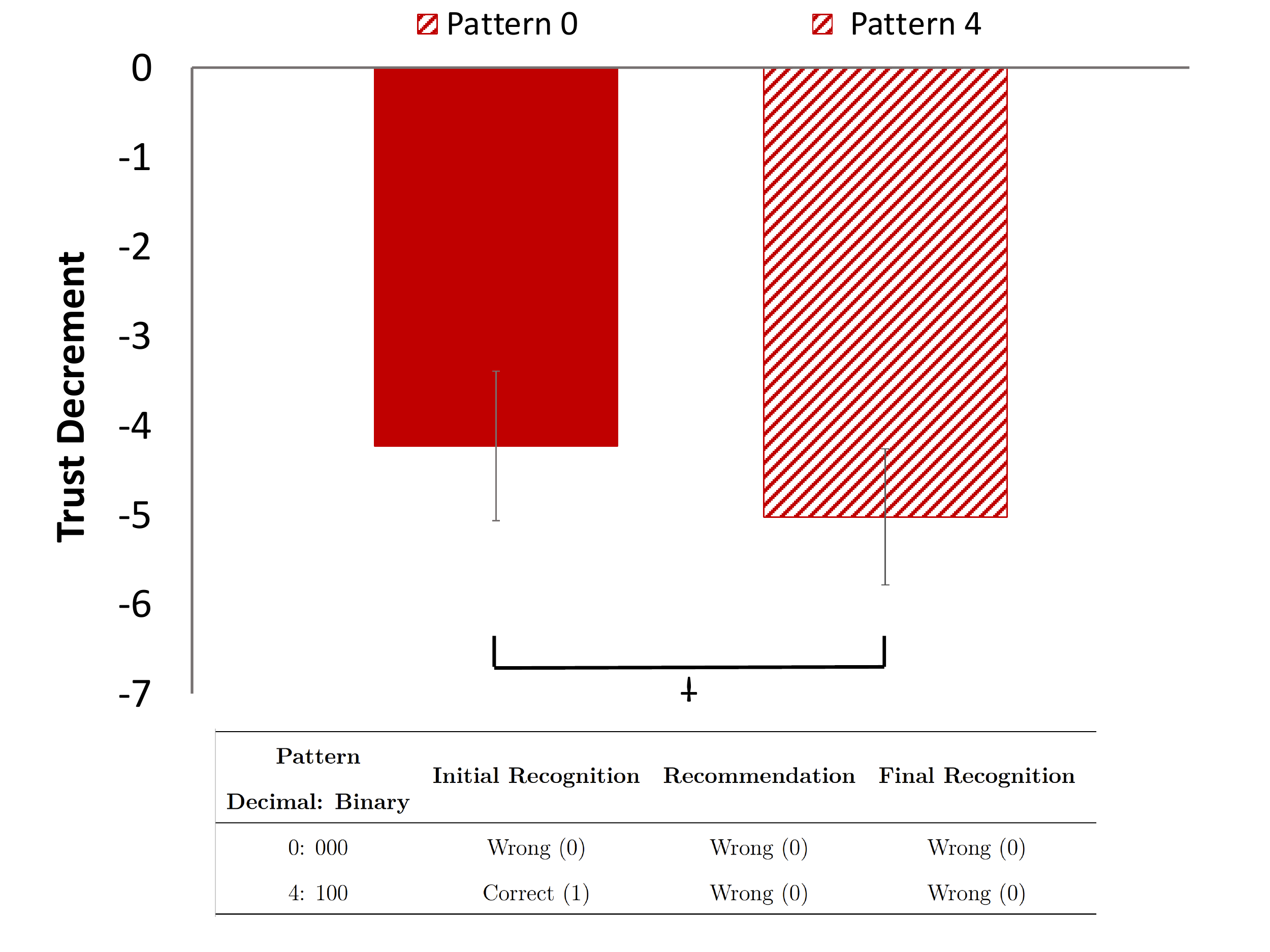}}

\caption{(a) Comparing \textit{\textbf{trust increment}} between patterns 3
and 7 (\textsuperscript{$\ast\ast\ast$}$p$  < .001).
(b) Comparing \textit{\textbf{trust decrement}} between patterns 0 
and 4 (\textsuperscript{$\dagger$}$p$  < .1 ).
}
\label{fig: contrast_effect}

\end{figure}

We hypothesize in \textbf{\textit{H3a}} that an automation success would produce a greater trust increment if the human operator fails the task.
We compare pattern 3 (wrong initial recognition - correct recommendation - correct final recognition) versus pattern 7 (correct initial recognition - correct recommendation - correct final recognition). In the two patterns, the automated aid provides correct recommendations, and the only difference is the human operators' initial recognition. A paired-samples t-test reveals a significantly larger trust increment in pattern 3 compared to pattern 7 ($t(1,69) = 4.40, p < 0.001$, Cohen's $d = 0.53$) (Figure \ref{fig: contrast_effect}(a)).





In \textbf{\textit{H3b}}, we hypothesize that an automation failure would lead to a greater trust decrement if the human operator is capable of performing the task.
We compare pattern 0 (wrong initial recognition - wrong recommendation - wrong final recognition) versus pattern 4 (correct initial recognition - wrong recommendation - wrong final recognition).  The analysis show a marginally significant difference between the two patterns ($t(1,57) = 1.83, p = .07$, Cohen's $d = 0.24$) (Figure \ref{fig: contrast_effect}(b)). 

After examining the moment-to-moment trust adjustments, we further explore how the accumulation of moment-to-moment adjustments contribute to the participants' trust at the end of the experiment. Table \ref{tlb:trust_pre_post} summarizes participants' trust propensity, trust after the 40th trial (i.e., after the entire interaction experience with automation), and post-experiment trust at each automation reliability level. 

\begin{table}[H]
\caption{Mean and standard deviation values of participants' trust propensity, $Trust(40)$, and post-experiment trust}
\begin{adjustbox}{width=1\textwidth}
\begin{tabular}{ccccc}
\hline
\textbf{Reliability (\%)} & \textbf{\begin{tabular}[c]{@{}c@{}}Trust Propensity \\ \end{tabular}} & \textbf{\begin{tabular}[c]{@{}c@{}}Trust after 40th Trial\\ $Trust(40)$\\ \end{tabular}} &
\textbf{\begin{tabular}[c]{@{}c@{}}Post-experiment Trust \\ Scale of \cite{Jian:2000es}\\ \end{tabular}} &
\textbf{\begin{tabular}[c]{@{}c@{}} Post-experiment Trust \\ Scale of \cite{MORAY:1996it} \end{tabular}}   \\ \hline
70 & 72.6 (14.8) & 56.6 (25.7) & 50.6 (18.0) & 53.3 (16.9) \\
80 & 69.4 (10.4) & 73.8 (17.0) & 54.1 (12.1)& 55.9 (12.4) \\
90 & 69.4 (14.4) & 81.6 (15.0) & 64.9 (16.7)& 72.1 (13.5)\\ \hline
\end{tabular}
\end{adjustbox}
\label{tlb:trust_pre_post}
\end{table}

We construct four linear regression models to explore how automation reliability affect trust propensity, and how reliability and trust propensity jointly influence participants' trust after they interact with the automation. The results show that automation reliability is not a significant predictor of trust propensity ($F(1, 73) = .73, p = .40$). Both automation reliability ($\beta = 1.32, t (72) = 4.91 , p <. 001$) and trust propensity ($ \beta = 0.45, t(72) = 2.72, p <.01 $) significantly predict participants' trust after the 40th trial. Automation reliability ($\beta = 0.77, t(72) =3.60, p <. 001$) and trust propensity ($ \beta = 0.37, t(72)=2.74,  p <.01 $) significantly predict participants' post-experiment trust measured by the scale of \citet{Jian:2000es}. Automation reliability ($\beta = 1.00, t(72) =5.10, p <. 001$) and trust propensity ($ \beta = 0.39, t(72)=3.16,  p <.01 $) significantly predict participants' post-experiment trust measured by the scale of \citet{MORAY:1996it}. 

\section{4. DISCUSSION}

Consistent with previous work of \citet{Lee:1992it}, \citet{Moray:2000} and \citet{Yang:2017:EEU:2909824.3020230, Yang:2016}, we also find that trust in automation increases as a result of automation successes and decreases as a result of automation failures (supporting \textbf{\textit{H1a}}). Our finding that the magnitude of trust decrements is larger than that of trust increments (supporting \textbf{\textit{H1b}}) is in line with previous studies 
in which the number of automation successes significantly exceeded that of automation failures \citep{manzey2012human, Lee:1992it}. In \citet{Lee:1992it}, participants experienced two automation failures among 50 experimental trials. In \citet{manzey2012human}, participants had one or two automation failure(s) among 20 automation successes. Both studies consistently showed that the strength of automation failures on trust adjustment is considerably stronger than automation successes. Our study show that the stronger effect of automation failures still holds when there is a considerable number of automation failures (i.e., 12 failures among 40 trials).


With respect to the outcome bias, we find supporting evidence for \textbf{\textit{H2b}} that an automation failure leads to a larger trust decrement if the final outcome (i.e., the final recognition performance) is undesirable. In line with previous research in medical, gambling and business decisions \citep{baron1988outcome, brownback2019understanding, gino2009see, sezer2016overcoming}, human operators demonstrate outcome biases in trust adjustments when the automation is wrong: an automation error is forgiven to some extent if the error does not lead to detrimental outcomes. 
This finding is disconcerting because the final outcome of a task can be due to a combination of factors. The automation influences but do not entirely determine the outcome of the task. This finding should be explored further in follow-up experiments with more cognitively demanding tasks, such as medical decision-making, as the experimental task used in the present study is fairly simple. 


However, results of the present study does not support \textbf{\textit{H2a}} that a correct recommendation would lead to a larger trust increment if the final outcome (i.e., the final recognition performance) is good. This non-significant result could have been due to the self-serving bias \citep{Weiner1985, Miller1975, Duval2002}. The self-serving bias is any cognitive or perceptual process that is distorted by the need to maintain and enhance self-esteem. It is particularly evident when individuals attribute the cause of outcomes. When explaining positive outcomes, their attributions emphasize the causal impact of internal, dispositional causes, but when identifying the causes of negative events, they stress external, situational factors. If we take a close look at the comparisons between pattern 2 (wrong initial recognition - correct recommendation - wrong final recognition) and pattern 3 (wrong initial recognition - correct recommendation - correct final recognition), the contrast between the two patterns suggest that the final correct outcome in Pattern 3 is largely due to the correct automation recommendation, and the final wrong outcome in Pattern 2 is largely due to the human operators' wrong initial recognition. According to the self-serving bias, the human operator would likely distort the trust adjustment process to maintain self-esteem, by appreciating the correct recommendation less than they should have done in Pattern 3.

 
Results of the present study also provide support for \textbf{\textit{H3a}} that an automation success produces a greater trust increment if the human operator fails the task and marginal support for \textbf{\textit{H3b}} that an automation failure produces a greater trust decrement if the human operator succeeds the task. 
The experimental paradigm used in the present study allows direct assessment of human operators' ability (i.e., manual performance without the help of automation aids). Instead of assessing the participants' ability directly, prior literature often evaluated participants' self-confidence in performing a task manually and considered operators' self-confidence and their trust in automation two independent constructs (i.e., Two constructs have no association). For example, \citet{DeVries2003} and \citet{lee1994trust} found that trust and self-confidence predict participants' automation dependence behaviors -- human operators use automation when trust exceeds self-confidence and use manual control when self-confidence exceeds trust. Our findings reveal that human operators' ability and their trust adjustment are not independent of each other. Because of the strong association between self-confidence and ability \citep{Wixted2017}, people's self-confidence and trust are probably not independent either, and therefore should not be viewed as independent constructs.


Viewing the results on \textbf{\textit{H1--H3}} holistically, our findings suggest that human operators are rational only to a certain extent when adjusting trust in automation. They are rational in the sense that they increase trust in automation after automation successes and decrease trust upon automation failures. On the contrary, they are irrational in the sense that their trust adjustments are significantly influenced by decision-making heuristics/biases. 

Along with the primary hypotheses, we are interested in exploring any potential impact of the overall automation reliability on trust adjustment. We find no significant effect on any of the one-way or pairwise t-tests (i.e., automation reliability does not significantly influence trust adjustment). On the contrary, we find that automation reliability significantly predicts the (snapshot) trust in automation measured at the end of an experiment, which is in line with previous research \citep{Du2019, DeVisser:2011dg, Wickens:2007hm}. Our findings suggest that effect of automation reliability on the (snapshot) trust at the end of an experiment is due to the accumulation of the moment-to-moment trust adjustments over time. A less reliable automation produces more automation failures, leading to a lower (snapshot) trust at the end.

\section{5. CONCLUSION}

In contrast to the snapshot view of trust, this study considers trust a dynamic variable and examines how human operators adjust their trust in automation as a result of moment-to-moment interaction with automation. Understanding the trust adjustment process enables accurate prediction of an operator's moment-to-moment trust in automation, which can be used to design trust-aware adaptive automation. 
To the best of our knowledge, it is the first to show that operators' trust adjustments are subject to decision-making heuristics/biases. It also provides further empirical evidence that automation failures have a greater impact on trust adjustments than automation successes. 

Moreover, we present a novel experimental paradigm that can be used to examine human operators' trust dynamics. The paradigm allows us to track participant's moment-to-moment trust over time. In addition, it distinguishes participants' ability to perform a task manually, the automated decision aid's performance, and the final performance. By eliciting human operators' answers pre- and post-automation recommendation, researchers could take a deep dive into how trust dynamics can be influenced by the interplay between participants' performance without the aid of automation, automation performance, and performance with the aid of automation.



We note the following limitations. Similar to a few previous studies \citep{manzey2012human, Yang:2017:EEU:2909824.3020230}, we used a one-item scale. It is possible, however, that one item will fail to capture all of the sub dimensions of trust compared to the use of multi-dimension scales such as the 12-item trust scale in \citet{Jian:2000es}.
Further research should investigate the possibility of a succinct multi-scale trust scale that can be used in querying trust over repeated interactions with automation. Second, the occurrence of patterns 1 and 6 was excluded from data analysis as their occurrences were rare. Further research could be conducted to purposely induce the occurrences of these two patterns.

\newpage

\section{Key points}
\begin{itemize}
\item Human operators adjust their trust in automation due to moment-to-moment interaction with automation. The trust adjustment process is moderated by decision-making heuristics/biases including outcome bias and contrast effect.

\item An automation failure is forgiven to a certain extent if the failure does not harm the final task outcome. 

\item An automation success engenders a larger trust increment if the human operator fails the task by him-/her-self. An automation failure leads to a marginally larger trust decrement if the human operator succeed the task. 

\item The stronger effect of automation failures on trust adjustment still holds when the occurrence of automation failures is up to 30\% (i.e., Automation reliability is 70\%). 

\end{itemize}
\newpage
\section{Appendix A}

\begin{table}[]
\begin{adjustbox}{width=1\textwidth}
\begin{tabular}{cl}
 & \multicolumn{1}{c}{Trust Propensity Survey (adapted from Merritt, Heimbaugh, LaChapell \& Deborah, 2013)}              \\ \hline
1    & I usually trust machines/automated technologies until there is a reason not to.                                        \\
2    & For the most part, I distrust machines/automated technologies.                                                         \\
3    & In general, I would rely on an automated machine/technology to assist me.                                              \\
4    & My tendency to trust machines/automated technologies is high.                                                          \\
5    & It is easy for me to trust machines/automated technologies to do their job.                                            \\
6    & I am likely to trust a machine/automated technology even when I have little knowledge about it.                        \\ \hline
\end{tabular}
\end{adjustbox}
\end{table}

\newpage
\section{Appendix B}
\begin{table}[]
\begin{adjustbox}{width=0.9\textwidth}
\begin{tabular}{cl}
 & \multicolumn{1}{c}{Post-experiment Trust Survey (adopted from Jian, Bisantz \& Drury, 2000)}              \\ \hline
1    & The automated decision aid is deceptive.                             \\
2    & The automated decision aid behaves in an underhanded manner.                     \\
3    & I am suspicious of the automated decision aid's intents, actions, or outputs.\\
4    & I am wary of the automated decision aid.                                     \\
5    & The automated decision aid's actions will have a harmful or injurious outcome.                                             \\
6    & I am confident in the automated decision aid.    \\
7    & he automated decision aid provides security. \\
8    & The automated decision aid has integrity.                                \\
9    & The automated decision aid is dependable.                                \\
10    & The automated decision aid is reliable.                                 \\
11    & I can trust the automated decision aid.                                     \\
12    & I am familiar with the automated decision aid.  
\\ \hline
\end{tabular}
\end{adjustbox}
\end{table}

\newpage
\section{Appendix C}
\begin{table}[]
\begin{adjustbox}{width=0.9\textwidth}
\begin{tabular}{cl}
 & \multicolumn{1}{c}{Post-experiment Trust Survey (adopted from Muir and Moray, 1996)}              \\ \hline
1    & To what extent does the automated decision aid perform its function properly?                            \\
2    & To what extent can the automated decision aid's behavior be predicted from moment to moment?                    \\
3    & To what extent can you count on the automated decision aid to do its job?    \\
4    & To what extent does the automated decision aid perform the task it was designed to do in the system?                         \\
5    & To what extent does the automated decision aid respond similarly to similar circumstances at different points in time?   \\
6    & My degree of faith that the automated decision aid will be able to cope with other system states in the future:   \\
7    & My degree of trust in the automated decision aid to respond accurately: \\
8    & My overall degree of trust in the automated decision aid:
\\ \hline
\end{tabular}
\end{adjustbox}

\textit{Note}: One item,``My degree of trust in the automated decision aid's display'' from the original survey was not included.
\end{table}

\bibliography{HFES-bibliography}

\begin{thebibliography}{}

\bibitem [\protect \citeauthoryear {%
Baron%
\ \BBA {} Hershey%
}{%
Baron%
\ \BBA {} Hershey%
}{%
{\protect \APACyear {1988}}%
}]{%
baron1988outcome}
\APACinsertmetastar {%
baron1988outcome}%
\begin{APACrefauthors}%
Baron, J.%
\BCBT {}\ \BBA {} Hershey, J\BPBI C.%
\end{APACrefauthors}%
\unskip\
\newblock
\APACrefYearMonthDay{1988}{}{}.
\newblock
{\BBOQ}\APACrefatitle {Outcome bias in decision evaluation.} {Outcome bias in
  decision evaluation.}{\BBCQ}
\newblock
\APACjournalVolNumPages{Journal of Personality and Social
  Psychology}{54}{4}{569}.
\newblock
\begin{APACrefDOI} \doi{https://doi.org/10.1037/0022-3514.54.4.569}
  \end{APACrefDOI}
\PrintBackRefs{\CurrentBib}

\bibitem [\protect \citeauthoryear {%
Bernstein%
, Wilson%
, Pernat%
\BCBL {}\ \BBA {} Meilleur%
}{%
Bernstein%
\ \protect \BOthers {.}}{%
{\protect \APACyear {2012}}%
}]{%
bernstein2012auditory}
\APACinsertmetastar {%
bernstein2012auditory}%
\begin{APACrefauthors}%
Bernstein, D\BPBI M.%
, Wilson, A\BPBI M.%
, Pernat, N\BPBI L.%
\BCBL {}\ \BBA {} Meilleur, L\BPBI R.%
\end{APACrefauthors}%
\unskip\
\newblock
\APACrefYearMonthDay{2012}{}{}.
\newblock
{\BBOQ}\APACrefatitle {Auditory hindsight bias} {Auditory hindsight
  bias}.{\BBCQ}
\newblock
\APACjournalVolNumPages{Psychonomic Bulletin \& Review}{19}{4}{588--593}.
\newblock
\begin{APACrefDOI} \doi{https://doi.org/10.3758/s13423-012-0268-0}
  \end{APACrefDOI}
\PrintBackRefs{\CurrentBib}

\bibitem [\protect \citeauthoryear {%
Brownback%
\ \BBA {} Kuhn%
}{%
Brownback%
\ \BBA {} Kuhn%
}{%
{\protect \APACyear {2019}}%
}]{%
brownback2019understanding}
\APACinsertmetastar {%
brownback2019understanding}%
\begin{APACrefauthors}%
Brownback, A.%
\BCBT {}\ \BBA {} Kuhn, M\BPBI A.%
\end{APACrefauthors}%
\unskip\
\newblock
\APACrefYearMonthDay{2019}{}{}.
\newblock
{\BBOQ}\APACrefatitle {Understanding outcome bias} {Understanding outcome
  bias}.{\BBCQ}
\newblock
\APACjournalVolNumPages{Games and Economic Behavior}{117}{}{342--360}.
\newblock
\begin{APACrefDOI} \doi{https://doi.org/10.1016/j.geb.2019.07.003}
  \end{APACrefDOI}
\PrintBackRefs{\CurrentBib}

\bibitem [\protect \citeauthoryear {%
de Visser%
\ \BBA {} Parasuraman%
}{%
de Visser%
\ \BBA {} Parasuraman%
}{%
{\protect \APACyear {2011}}%
}]{%
DeVisser:2011dg}
\APACinsertmetastar {%
DeVisser:2011dg}%
\begin{APACrefauthors}%
de Visser, E\BPBI J.%
\BCBT {}\ \BBA {} Parasuraman, R.%
\end{APACrefauthors}%
\unskip\
\newblock
\APACrefYearMonthDay{2011}{}{}.
\newblock
{\BBOQ}\APACrefatitle {Adaptive Aiding of Human-Robot Teaming} {Adaptive aiding
  of human-robot teaming}.{\BBCQ}
\newblock
\APACjournalVolNumPages{Journal of Cognitive Engineering and Decision
  Making}{5}{2}{209--231}.
\newblock
\begin{APACrefDOI} \doi{https://doi.org/10.1177/1555343411410160}
  \end{APACrefDOI}
\PrintBackRefs{\CurrentBib}

\bibitem [\protect \citeauthoryear {%
de Visser%
\ \protect \BOthers {.}}{%
de Visser%
\ \protect \BOthers {.}}{%
{\protect \APACyear {2020}}%
}]{%
DeVisser_IJSR}
\APACinsertmetastar {%
DeVisser_IJSR}%
\begin{APACrefauthors}%
de Visser, E\BPBI J.%
, Peeters, M\BPBI M.%
, Jung, M\BPBI F.%
, Kohn, S.%
, Shaw, T\BPBI H.%
, Pak, R.%
\BCBL {}\ \BBA {} Neerincx, M\BPBI A.%
\end{APACrefauthors}%
\unskip\
\newblock
\APACrefYearMonthDay{2020}{}{}.
\newblock
{\BBOQ}\APACrefatitle {Towards a Theory of Longitudinal Trust Calibration in
  Human–Robot Teams} {Towards a theory of longitudinal trust calibration in
  human–robot teams}.{\BBCQ}
\newblock
\APACjournalVolNumPages{International Journal of Social
  Robotics}{12}{2}{459--478}.
\newblock
\begin{APACrefDOI} \doi{https://doi.org/10.1007/s12369-019-00596-x}
  \end{APACrefDOI}
\PrintBackRefs{\CurrentBib}

\bibitem [\protect \citeauthoryear {%
de Vries%
, Midden%
\BCBL {}\ \BBA {} Bouwhuis%
}{%
de Vries%
\ \protect \BOthers {.}}{%
{\protect \APACyear {2003}}%
}]{%
DeVries2003}
\APACinsertmetastar {%
DeVries2003}%
\begin{APACrefauthors}%
de Vries, P.%
, Midden, C.%
\BCBL {}\ \BBA {} Bouwhuis, D.%
\end{APACrefauthors}%
\unskip\
\newblock
\APACrefYearMonthDay{2003}{}{}.
\newblock
{\BBOQ}\APACrefatitle {The effects of errors on system trust, self-confidence,
  and the allocation of control in route planning} {The effects of errors on
  system trust, self-confidence, and the allocation of control in route
  planning}.{\BBCQ}
\newblock
\APACjournalVolNumPages{International Journal of Human-Computer
  Studies}{58}{6}{719--735}.
\newblock
\begin{APACrefDOI} \doi{https://doi.org/10.1016/S1071-5819(03)00039-9}
  \end{APACrefDOI}
\PrintBackRefs{\CurrentBib}

\bibitem [\protect \citeauthoryear {%
Du%
, Huang%
\BCBL {}\ \BBA {} Yang%
}{%
Du%
\ \protect \BOthers {.}}{%
{\protect \APACyear {2020}}%
}]{%
Du2019}
\APACinsertmetastar {%
Du2019}%
\begin{APACrefauthors}%
Du, N.%
, Huang, K\BPBI Y.%
\BCBL {}\ \BBA {} Yang, X\BPBI J.%
\end{APACrefauthors}%
\unskip\
\newblock
\APACrefYearMonthDay{2020}{}{}.
\newblock
{\BBOQ}\APACrefatitle {Not All Information Is Equal: Effects of Disclosing
  Different Types of Likelihood Information on Trust, Compliance and Reliance,
  and Task Performance in Human-Automation Teaming} {Not all information is
  equal: Effects of disclosing different types of likelihood information on
  trust, compliance and reliance, and task performance in human-automation
  teaming}.{\BBCQ}
\newblock
\APACjournalVolNumPages{Human Factors}{62}{6}{987-1001}.
\newblock
\begin{APACrefDOI} \doi{https://doi.org/10.1177/0018720819862916}
  \end{APACrefDOI}
\PrintBackRefs{\CurrentBib}

\bibitem [\protect \citeauthoryear {%
Duval%
\ \BBA {} Silvia%
}{%
Duval%
\ \BBA {} Silvia%
}{%
{\protect \APACyear {2002}}%
}]{%
Duval2002}
\APACinsertmetastar {%
Duval2002}%
\begin{APACrefauthors}%
Duval, T\BPBI S.%
\BCBT {}\ \BBA {} Silvia, P\BPBI J.%
\end{APACrefauthors}%
\unskip\
\newblock
\APACrefYearMonthDay{2002}{}{}.
\newblock
{\BBOQ}\APACrefatitle {{Self-awareness, probability of improvement, and the
  self-serving bias}} {{Self-awareness, probability of improvement, and the
  self-serving bias}}.{\BBCQ}
\newblock
\APACjournalVolNumPages{Journal of Personality and Social
  Psychology}{82}{1}{49--61}.
\newblock
\begin{APACrefDOI} \doi{https://10.1037/0022-3514.82.1.49} \end{APACrefDOI}
\PrintBackRefs{\CurrentBib}

\bibitem [\protect \citeauthoryear {%
Fischhoff%
}{%
Fischhoff%
}{%
{\protect \APACyear {1975}}%
}]{%
fischhoff1975hindsight}
\APACinsertmetastar {%
fischhoff1975hindsight}%
\begin{APACrefauthors}%
Fischhoff, B.%
\end{APACrefauthors}%
\unskip\
\newblock
\APACrefYearMonthDay{1975}{}{}.
\newblock
{\BBOQ}\APACrefatitle {Hindsight is not equal to foresight: The effect of
  outcome knowledge on judgment under uncertainty} {Hindsight is not equal to
  foresight: The effect of outcome knowledge on judgment under
  uncertainty}.{\BBCQ}
\newblock
\APACjournalVolNumPages{Journal of Experimental Psychology: Human Perception
  and Performance}{1}{3}{288}.
\newblock
\begin{APACrefDOI} \doi{https://doi.org/10.1037/0096-1523.1.3.288}
  \end{APACrefDOI}
\PrintBackRefs{\CurrentBib}

\bibitem [\protect \citeauthoryear {%
Gino%
, Moore%
\BCBL {}\ \BBA {} Bazerman%
}{%
Gino%
\ \protect \BOthers {.}}{%
{\protect \APACyear {2009}}%
}]{%
gino2009see}
\APACinsertmetastar {%
gino2009see}%
\begin{APACrefauthors}%
Gino, F.%
, Moore, D\BPBI A.%
\BCBL {}\ \BBA {} Bazerman, M\BPBI H.%
\end{APACrefauthors}%
\unskip\
\newblock
\APACrefYearMonthDay{2009}{}{}.
\newblock
{\BBOQ}\APACrefatitle {See no evil: When we overlook other people’s unethical
  behavior} {See no evil: When we overlook other people’s unethical
  behavior}.{\BBCQ}
\newblock
\APACjournalVolNumPages{Social decision making: Social Dilemmas, Social Values,
  and Ethical Judgments}{}{}{241--263}.
\newblock
\begin{APACrefDOI} \doi{https://doi.org/10.1037/0096-1523.1.3.288}
  \end{APACrefDOI}
\PrintBackRefs{\CurrentBib}

\bibitem [\protect \citeauthoryear {%
Guilbault%
, Bryant%
, Brockway%
\BCBL {}\ \BBA {} Posavac%
}{%
Guilbault%
\ \protect \BOthers {.}}{%
{\protect \APACyear {2004}}%
}]{%
guilbault2004meta}
\APACinsertmetastar {%
guilbault2004meta}%
\begin{APACrefauthors}%
Guilbault, R\BPBI L.%
, Bryant, F\BPBI B.%
, Brockway, J\BPBI H.%
\BCBL {}\ \BBA {} Posavac, E\BPBI J.%
\end{APACrefauthors}%
\unskip\
\newblock
\APACrefYearMonthDay{2004}{}{}.
\newblock
{\BBOQ}\APACrefatitle {A meta-analysis of research on hindsight bias} {A
  meta-analysis of research on hindsight bias}.{\BBCQ}
\newblock
\APACjournalVolNumPages{Basic and Applied Social
  Psychology}{26}{2-3}{103--117}.
\newblock
\begin{APACrefDOI} \doi{https://doi.org/10.1207/s15324834basp2602&3_1}
  \end{APACrefDOI}
\PrintBackRefs{\CurrentBib}

\bibitem [\protect \citeauthoryear {%
Guo%
, Shi%
\BCBL {}\ \BBA {} Yang%
}{%
Guo%
\ \protect \BOthers {.}}{%
{\protect \APACyear {2021}}%
}]{%
Guo2021_RAL}
\APACinsertmetastar {%
Guo2021_RAL}%
\begin{APACrefauthors}%
Guo, Y.%
, Shi, C.%
\BCBL {}\ \BBA {} Yang, X\BPBI J.%
\end{APACrefauthors}%
\unskip\
\newblock
\APACrefYearMonthDay{2021}{}{}.
\newblock
{\BBOQ}\APACrefatitle {{Reverse Psychology in Trust-Aware Human-Robot
  Interaction}} {{Reverse Psychology in Trust-Aware Human-Robot
  Interaction}}.{\BBCQ}
\newblock
\APACjournalVolNumPages{IEEE Robotics and Automation
  Letters}{6}{3}{4851--4858}.
\newblock
\begin{APACrefDOI} \doi{https://doi.org/10.1109/LRA.2021.3067626}
  \end{APACrefDOI}
\PrintBackRefs{\CurrentBib}

\bibitem [\protect \citeauthoryear {%
Guo%
\ \BBA {} Yang%
}{%
Guo%
\ \BBA {} Yang%
}{%
{\protect \APACyear {2020}}%
}]{%
Guo2020_IJSR}
\APACinsertmetastar {%
Guo2020_IJSR}%
\begin{APACrefauthors}%
Guo, Y.%
\BCBT {}\ \BBA {} Yang, X\BPBI J.%
\end{APACrefauthors}%
\unskip\
\newblock
\APACrefYearMonthDay{2020}{}{}.
\newblock
{\BBOQ}\APACrefatitle {Modeling and Predicting Trust Dynamics in Human–Robot
  Teaming: A Bayesian Inference Approach} {Modeling and predicting trust
  dynamics in human–robot teaming: A bayesian inference approach}.{\BBCQ}
\newblock
\APACjournalVolNumPages{International Journal of Social Robotics}{}{}{}.
\newblock
\begin{APACrefDOI} \doi{https://doi.org/10.1007/s12369-020-00703-3}
  \end{APACrefDOI}
\PrintBackRefs{\CurrentBib}

\bibitem [\protect \citeauthoryear {%
Hawkins%
\ \BBA {} Hastie%
}{%
Hawkins%
\ \BBA {} Hastie%
}{%
{\protect \APACyear {1990}}%
}]{%
hawkins1990hindsight}
\APACinsertmetastar {%
hawkins1990hindsight}%
\begin{APACrefauthors}%
Hawkins, S\BPBI A.%
\BCBT {}\ \BBA {} Hastie, R.%
\end{APACrefauthors}%
\unskip\
\newblock
\APACrefYearMonthDay{1990}{}{}.
\newblock
{\BBOQ}\APACrefatitle {Hindsight: Biased judgments of past events after the
  outcomes are known.} {Hindsight: Biased judgments of past events after the
  outcomes are known.}{\BBCQ}
\newblock
\APACjournalVolNumPages{Psychological Bulletin}{107}{3}{311}.
\newblock
\begin{APACrefDOI} \doi{https://doi.org/10.1037/0033-2909.107.3.311}
  \end{APACrefDOI}
\PrintBackRefs{\CurrentBib}

\bibitem [\protect \citeauthoryear {%
Henriksen%
\ \BBA {} Kaplan%
}{%
Henriksen%
\ \BBA {} Kaplan%
}{%
{\protect \APACyear {2003}}%
}]{%
henriksen2003hindsight}
\APACinsertmetastar {%
henriksen2003hindsight}%
\begin{APACrefauthors}%
Henriksen, K.%
\BCBT {}\ \BBA {} Kaplan, H.%
\end{APACrefauthors}%
\unskip\
\newblock
\APACrefYearMonthDay{2003}{}{}.
\newblock
{\BBOQ}\APACrefatitle {{Hindsight bias, outcome knowledge and adaptive
  learning}} {{Hindsight bias, outcome knowledge and adaptive
  learning}}.{\BBCQ}
\newblock
\APACjournalVolNumPages{BMJ Quality \& Safety}{12}{suppl 2}{ii46----ii50}.
\newblock
\begin{APACrefDOI} \doi{https://doi.org/10.1136/qhc.12.suppl_2.ii46}
  \end{APACrefDOI}
\PrintBackRefs{\CurrentBib}

\bibitem [\protect \citeauthoryear {%
Hockey%
, Wastell%
\BCBL {}\ \BBA {} Sauer%
}{%
Hockey%
\ \protect \BOthers {.}}{%
{\protect \APACyear {1998}}%
}]{%
Hockey1998}
\APACinsertmetastar {%
Hockey1998}%
\begin{APACrefauthors}%
Hockey, G\BPBI R\BPBI J.%
, Wastell, D\BPBI G.%
\BCBL {}\ \BBA {} Sauer, J.%
\end{APACrefauthors}%
\unskip\
\newblock
\APACrefYearMonthDay{1998}{}{}.
\newblock
{\BBOQ}\APACrefatitle {{Effects of sleep deprivation and user interface on
  complex performance: A multilevel analysis of compensatory control}}
  {{Effects of sleep deprivation and user interface on complex performance: A
  multilevel analysis of compensatory control}}.{\BBCQ}
\newblock
\APACjournalVolNumPages{Human Factors}{40}{2}{233--253}.
\newblock
\begin{APACrefDOI} \doi{https://doi.org/10.1518/001872098779480479}
  \end{APACrefDOI}
\PrintBackRefs{\CurrentBib}

\bibitem [\protect \citeauthoryear {%
Hsu%
\ \BBA {} Lee%
}{%
Hsu%
\ \BBA {} Lee%
}{%
{\protect \APACyear {2016}}%
}]{%
hsu2016relative}
\APACinsertmetastar {%
hsu2016relative}%
\begin{APACrefauthors}%
Hsu, S\BHBI M.%
\BCBT {}\ \BBA {} Lee, J\BHBI S.%
\end{APACrefauthors}%
\unskip\
\newblock
\APACrefYearMonthDay{2016}{}{}.
\newblock
{\BBOQ}\APACrefatitle {Relative judgment in facial identity perception as
  revealed by sequential effects} {Relative judgment in facial identity
  perception as revealed by sequential effects}.{\BBCQ}
\newblock
\APACjournalVolNumPages{Attention, Perception, \&
  Psychophysics}{78}{1}{264--277}.
\newblock
\begin{APACrefDOI} \doi{https://doi.org/10.3758/s13414-015-0979-1}
  \end{APACrefDOI}
\PrintBackRefs{\CurrentBib}

\bibitem [\protect \citeauthoryear {%
Jian%
, Bisantz%
\BCBL {}\ \BBA {} Drury%
}{%
Jian%
\ \protect \BOthers {.}}{%
{\protect \APACyear {{\protect \bibnodate {}}}}%
}]{%
Jian:2000es}
\APACinsertmetastar {%
Jian:2000es}%
\begin{APACrefauthors}%
Jian, J\BHBI Y.%
, Bisantz, A\BPBI M.%
\BCBL {}\ \BBA {} Drury, C\BPBI G.%
\end{APACrefauthors}%
\unskip\
\newblock
\APACrefYearMonthDay{{\protect \bibnodate {}}}{}{}.
\newblock
{\BBOQ}\APACrefatitle {Foundations for an Empirically Determined Scale of Trust
  in Automated Systems} {Foundations for an empirically determined scale of
  trust in automated systems}.{\BBCQ}
\newblock
\APACjournalVolNumPages{International Journal of Cognitive
  Ergonomics}{4}{1}{53--71}.
\newblock
\begin{APACrefDOI} \doi{https://doi.org10.1207/S15327566IJCE0401_04}
  \end{APACrefDOI}
\PrintBackRefs{\CurrentBib}

\bibitem [\protect \citeauthoryear {%
Kausel%
, Ventura%
\BCBL {}\ \BBA {} Rodr{\'\i}guez%
}{%
Kausel%
\ \protect \BOthers {.}}{%
{\protect \APACyear {2019}}%
}]{%
kausel2019outcome}
\APACinsertmetastar {%
kausel2019outcome}%
\begin{APACrefauthors}%
Kausel, E\BPBI E.%
, Ventura, S.%
\BCBL {}\ \BBA {} Rodr{\'\i}guez, A.%
\end{APACrefauthors}%
\unskip\
\newblock
\APACrefYearMonthDay{2019}{}{}.
\newblock
{\BBOQ}\APACrefatitle {Outcome bias in subjective ratings of performance:
  Evidence from the (football) field} {Outcome bias in subjective ratings of
  performance: Evidence from the (football) field}.{\BBCQ}
\newblock
\APACjournalVolNumPages{Journal of Economic Psychology}{75}{}{102132}.
\newblock
\begin{APACrefDOI} \doi{https://doi.org/10.1016/j.joep.2018.12.006}
  \end{APACrefDOI}
\PrintBackRefs{\CurrentBib}

\bibitem [\protect \citeauthoryear {%
Kenrick%
\ \BBA {} Gutierres%
}{%
Kenrick%
\ \BBA {} Gutierres%
}{%
{\protect \APACyear {1980}}%
}]{%
kenrick1980contrast}
\APACinsertmetastar {%
kenrick1980contrast}%
\begin{APACrefauthors}%
Kenrick, D\BPBI T.%
\BCBT {}\ \BBA {} Gutierres, S\BPBI E.%
\end{APACrefauthors}%
\unskip\
\newblock
\APACrefYearMonthDay{1980}{}{}.
\newblock
{\BBOQ}\APACrefatitle {Contrast effects and judgments of physical
  attractiveness: When beauty becomes a social problem.} {Contrast effects and
  judgments of physical attractiveness: When beauty becomes a social
  problem.}{\BBCQ}
\newblock
\APACjournalVolNumPages{Journal of Personality and Social
  Psychology}{38}{1}{131}.
\newblock
\begin{APACrefDOI} \doi{https://doi.org/10.1037/0022-3514.38.1.131}
  \end{APACrefDOI}
\PrintBackRefs{\CurrentBib}

\bibitem [\protect \citeauthoryear {%
Lee%
\ \BBA {} Moray%
}{%
Lee%
\ \BBA {} Moray%
}{%
{\protect \APACyear {1992}}%
}]{%
Lee:1992it}
\APACinsertmetastar {%
Lee:1992it}%
\begin{APACrefauthors}%
Lee, J\BPBI D.%
\BCBT {}\ \BBA {} Moray, N.%
\end{APACrefauthors}%
\unskip\
\newblock
\APACrefYearMonthDay{1992}{}{}.
\newblock
{\BBOQ}\APACrefatitle {Trust, control strategies and allocation of function in
  human-machine systems} {Trust, control strategies and allocation of function
  in human-machine systems}.{\BBCQ}
\newblock
\APACjournalVolNumPages{Ergonomics}{35}{10}{1243--1270}.
\newblock
\begin{APACrefDOI} \doi{https://doi.org/10.1080/00140139208967392}
  \end{APACrefDOI}
\PrintBackRefs{\CurrentBib}

\bibitem [\protect \citeauthoryear {%
Lee%
\ \BBA {} Moray%
}{%
Lee%
\ \BBA {} Moray%
}{%
{\protect \APACyear {1994}}%
}]{%
lee1994trust}
\APACinsertmetastar {%
lee1994trust}%
\begin{APACrefauthors}%
Lee, J\BPBI D.%
\BCBT {}\ \BBA {} Moray, N.%
\end{APACrefauthors}%
\unskip\
\newblock
\APACrefYearMonthDay{1994}{}{}.
\newblock
{\BBOQ}\APACrefatitle {Trust, self-confidence, and operators' adaptation to
  automation} {Trust, self-confidence, and operators' adaptation to
  automation}.{\BBCQ}
\newblock
\APACjournalVolNumPages{International Journal of Human-Computer
  Studies}{40}{1}{153--184}.
\newblock
\begin{APACrefDOI} \doi{https://doi.org/10.1006/ijhc.1994.1007}
  \end{APACrefDOI}
\PrintBackRefs{\CurrentBib}

\bibitem [\protect \citeauthoryear {%
Lynch%
, Chakravarti%
\BCBL {}\ \BBA {} Mitra%
}{%
Lynch%
\ \protect \BOthers {.}}{%
{\protect \APACyear {1991}}%
}]{%
lynch1991contrast}
\APACinsertmetastar {%
lynch1991contrast}%
\begin{APACrefauthors}%
Lynch, J\BPBI G., Jr.%
, Chakravarti, D.%
\BCBL {}\ \BBA {} Mitra, A.%
\end{APACrefauthors}%
\unskip\
\newblock
\APACrefYearMonthDay{1991}{}{}.
\newblock
{\BBOQ}\APACrefatitle {Contrast effects in consumer judgments: Changes in
  mental representations or in the anchoring of rating scales?} {Contrast
  effects in consumer judgments: Changes in mental representations or in the
  anchoring of rating scales?}{\BBCQ}
\newblock
\APACjournalVolNumPages{Journal of Consumer Research}{18}{3}{284--297}.
\newblock
\begin{APACrefDOI} \doi{https://doi.org/10.1086/209260} \end{APACrefDOI}
\PrintBackRefs{\CurrentBib}

\bibitem [\protect \citeauthoryear {%
Manzey%
, Reichenbach%
\BCBL {}\ \BBA {} Onnasch%
}{%
Manzey%
\ \protect \BOthers {.}}{%
{\protect \APACyear {2012}}%
}]{%
manzey2012human}
\APACinsertmetastar {%
manzey2012human}%
\begin{APACrefauthors}%
Manzey, D.%
, Reichenbach, J.%
\BCBL {}\ \BBA {} Onnasch, L.%
\end{APACrefauthors}%
\unskip\
\newblock
\APACrefYearMonthDay{2012}{}{}.
\newblock
{\BBOQ}\APACrefatitle {Human performance consequences of automated decision
  aids: The impact of degree of automation and system experience} {Human
  performance consequences of automated decision aids: The impact of degree of
  automation and system experience}.{\BBCQ}
\newblock
\APACjournalVolNumPages{Journal of Cognitive Engineering and Decision
  Making}{6}{1}{57--87}.
\PrintBackRefs{\CurrentBib}

\bibitem [\protect \citeauthoryear {%
McBride%
, Rogers%
\BCBL {}\ \BBA {} Fisk%
}{%
McBride%
\ \protect \BOthers {.}}{%
{\protect \APACyear {2011}}%
}]{%
McBride:2011ix}
\APACinsertmetastar {%
McBride:2011ix}%
\begin{APACrefauthors}%
McBride, S\BPBI E.%
, Rogers, W\BPBI A.%
\BCBL {}\ \BBA {} Fisk, A\BPBI D.%
\end{APACrefauthors}%
\unskip\
\newblock
\APACrefYearMonthDay{2011}{}{}.
\newblock
{\BBOQ}\APACrefatitle {Understanding the effect of workload on automation use
  for younger and older adults} {Understanding the effect of workload on
  automation use for younger and older adults}.{\BBCQ}
\newblock
\APACjournalVolNumPages{Human Factors}{53}{6}{672--686}.
\newblock
\begin{APACrefDOI} \doi{https://doi.org/10.1177/0018720811421909}
  \end{APACrefDOI}
\PrintBackRefs{\CurrentBib}

\bibitem [\protect \citeauthoryear {%
Merritt%
, Heimbaugh%
, LaChapell%
\BCBL {}\ \BBA {} Deborah%
}{%
Merritt%
\ \protect \BOthers {.}}{%
{\protect \APACyear {2013}}%
}]{%
merritt2013}
\APACinsertmetastar {%
merritt2013}%
\begin{APACrefauthors}%
Merritt, S\BPBI M.%
, Heimbaugh, H.%
, LaChapell, J.%
\BCBL {}\ \BBA {} Deborah, L.%
\end{APACrefauthors}%
\unskip\
\newblock
\APACrefYearMonthDay{2013}{}{}.
\newblock
{\BBOQ}\APACrefatitle {I Trust It, but I Don’t Know Why: Effects of Implicit
  Attitudes Toward Automation on Trust in an Automated System} {I trust it, but
  i don’t know why: Effects of implicit attitudes toward automation on trust
  in an automated system}.{\BBCQ}
\newblock
\APACjournalVolNumPages{Human Factors}{55}{3}{520-534}.
\newblock
\begin{APACrefDOI} \doi{https://doi.org/10.1177/0018720812465081}
  \end{APACrefDOI}
\PrintBackRefs{\CurrentBib}

\bibitem [\protect \citeauthoryear {%
Merritt%
\ \BBA {} Ilgen%
}{%
Merritt%
\ \BBA {} Ilgen%
}{%
{\protect \APACyear {2008}}%
}]{%
Merritt:2008ds}
\APACinsertmetastar {%
Merritt:2008ds}%
\begin{APACrefauthors}%
Merritt, S\BPBI M.%
\BCBT {}\ \BBA {} Ilgen, D\BPBI R.%
\end{APACrefauthors}%
\unskip\
\newblock
\APACrefYearMonthDay{2008}{}{}.
\newblock
{\BBOQ}\APACrefatitle {Not All Trust Is Created Equal: Dispositional and
  History-Based Trust in Human-Automation Interactions} {Not all trust is
  created equal: Dispositional and history-based trust in human-automation
  interactions}.{\BBCQ}
\newblock
\APACjournalVolNumPages{Human Factors}{50}{2}{194--210}.
\newblock
\begin{APACrefDOI} \doi{https://doi.org/10.1518/001872008X288574}
  \end{APACrefDOI}
\PrintBackRefs{\CurrentBib}

\bibitem [\protect \citeauthoryear {%
Miller%
\ \BBA {} Ross%
}{%
Miller%
\ \BBA {} Ross%
}{%
{\protect \APACyear {1975}}%
}]{%
Miller1975}
\APACinsertmetastar {%
Miller1975}%
\begin{APACrefauthors}%
Miller, D\BPBI T.%
\BCBT {}\ \BBA {} Ross, M.%
\end{APACrefauthors}%
\unskip\
\newblock
\APACrefYearMonthDay{1975}{}{}.
\newblock
{\BBOQ}\APACrefatitle {{Self-serving biases in the attribution of causality:
  Fact or fiction?}} {{Self-serving biases in the attribution of causality:
  Fact or fiction?}}{\BBCQ}
\newblock
\APACjournalVolNumPages{Psychological Bulletin}{82}{2}{213--225}.
\newblock
\begin{APACrefDOI} \doi{https://doi.org/10.1037/h0076486} \end{APACrefDOI}
\PrintBackRefs{\CurrentBib}

\bibitem [\protect \citeauthoryear {%
Moray%
, Inagaki%
\BCBL {}\ \BBA {} Itoh%
}{%
Moray%
\ \protect \BOthers {.}}{%
{\protect \APACyear {2000}}%
}]{%
Moray:2000}
\APACinsertmetastar {%
Moray:2000}%
\begin{APACrefauthors}%
Moray, N.%
, Inagaki, T.%
\BCBL {}\ \BBA {} Itoh, M.%
\end{APACrefauthors}%
\unskip\
\newblock
\APACrefYearMonthDay{2000}{}{}.
\newblock
{\BBOQ}\APACrefatitle {{Adaptive automation, trust, and self-confidence in
  fault management of time-critical tasks}} {{Adaptive automation, trust, and
  self-confidence in fault management of time-critical tasks}}.{\BBCQ}
\newblock
\APACjournalVolNumPages{Journal of Experimental Psychology
  Applied}{6}{1}{44--58}.
\PrintBackRefs{\CurrentBib}

\bibitem [\protect \citeauthoryear {%
Muir%
\ \BBA {} Moray%
}{%
Muir%
\ \BBA {} Moray%
}{%
{\protect \APACyear {1996}}%
}]{%
MORAY:1996it}
\APACinsertmetastar {%
MORAY:1996it}%
\begin{APACrefauthors}%
Muir, B\BPBI M.%
\BCBT {}\ \BBA {} Moray, N.%
\end{APACrefauthors}%
\unskip\
\newblock
\APACrefYearMonthDay{1996}{}{}.
\newblock
{\BBOQ}\APACrefatitle {{Trust in automation. Part II. Experimental studies of
  trust and human intervention in a process control simulation}} {{Trust in
  automation. Part II. Experimental studies of trust and human intervention in
  a process control simulation}}.{\BBCQ}
\newblock
\APACjournalVolNumPages{Ergonomics}{39}{3}{429--460}.
\newblock
\begin{APACrefDOI} \doi{https://doi.org/10.1080/00140139608964474}
  \end{APACrefDOI}
\PrintBackRefs{\CurrentBib}

\bibitem [\protect \citeauthoryear {%
Parasuraman%
, Sheridan%
\BCBL {}\ \BBA {} Wickens%
}{%
Parasuraman%
\ \protect \BOthers {.}}{%
{\protect \APACyear {2000}}%
}]{%
Parasuraman:2000}
\APACinsertmetastar {%
Parasuraman:2000}%
\begin{APACrefauthors}%
Parasuraman, R.%
, Sheridan, T\BPBI B.%
\BCBL {}\ \BBA {} Wickens, C\BPBI D.%
\end{APACrefauthors}%
\unskip\
\newblock
\APACrefYearMonthDay{2000}{}{}.
\newblock
{\BBOQ}\APACrefatitle {A model for types and levels of human interaction with
  automation} {A model for types and levels of human interaction with
  automation}.{\BBCQ}
\newblock
\APACjournalVolNumPages{IEEE Transactions on Systems, Man, and Cybernetics -
  Part A: Systems and Humans}{30}{3}{286--297}.
\newblock
\begin{APACrefDOI} \doi{https://doi.org/10.1109/3468.844354} \end{APACrefDOI}
\PrintBackRefs{\CurrentBib}

\bibitem [\protect \citeauthoryear {%
Rau%
, Li%
\BCBL {}\ \BBA {} Li%
}{%
Rau%
\ \protect \BOthers {.}}{%
{\protect \APACyear {2009}}%
}]{%
Rau2009}
\APACinsertmetastar {%
Rau2009}%
\begin{APACrefauthors}%
Rau, P\BPBI P.%
, Li, Y.%
\BCBL {}\ \BBA {} Li, D.%
\end{APACrefauthors}%
\unskip\
\newblock
\APACrefYearMonthDay{2009}{}{}.
\newblock
{\BBOQ}\APACrefatitle {Effects of communication style and culture on ability to
  accept recommendations from robots} {Effects of communication style and
  culture on ability to accept recommendations from robots}.{\BBCQ}
\newblock
\APACjournalVolNumPages{Computers in Human Behavior}{25}{2}{587--595}.
\newblock
\begin{APACrefDOI} \doi{https://doi.org/10.1016/j.chb.2008.12.025}
  \end{APACrefDOI}
\PrintBackRefs{\CurrentBib}

\bibitem [\protect \citeauthoryear {%
Roese%
\ \BBA {} Vohs%
}{%
Roese%
\ \BBA {} Vohs%
}{%
{\protect \APACyear {2012}}%
}]{%
roese2012hindsight}
\APACinsertmetastar {%
roese2012hindsight}%
\begin{APACrefauthors}%
Roese, N\BPBI J.%
\BCBT {}\ \BBA {} Vohs, K\BPBI D.%
\end{APACrefauthors}%
\unskip\
\newblock
\APACrefYearMonthDay{2012}{}{}.
\newblock
{\BBOQ}\APACrefatitle {Hindsight bias} {Hindsight bias}.{\BBCQ}
\newblock
\APACjournalVolNumPages{{Perspectives on Psychological
  Science}}{7}{5}{411--426}.
\newblock
\begin{APACrefDOI} \doi{https://doi.org/10.1177/1745691612454303}
  \end{APACrefDOI}
\PrintBackRefs{\CurrentBib}

\bibitem [\protect \citeauthoryear {%
Sezer%
, Zhang%
, Gino%
\BCBL {}\ \BBA {} Bazerman%
}{%
Sezer%
\ \protect \BOthers {.}}{%
{\protect \APACyear {2016}}%
}]{%
sezer2016overcoming}
\APACinsertmetastar {%
sezer2016overcoming}%
\begin{APACrefauthors}%
Sezer, O.%
, Zhang, T.%
, Gino, F.%
\BCBL {}\ \BBA {} Bazerman, M\BPBI H.%
\end{APACrefauthors}%
\unskip\
\newblock
\APACrefYearMonthDay{2016}{}{}.
\newblock
{\BBOQ}\APACrefatitle {Overcoming the outcome bias: Making intentions matter}
  {Overcoming the outcome bias: Making intentions matter}.{\BBCQ}
\newblock
\APACjournalVolNumPages{Organizational Behavior and Human Decision
  Processes}{137}{}{13--26}.
\newblock
\begin{APACrefDOI} \doi{https://doi.org/10.1016/j.obhdp.2016.07.001}
  \end{APACrefDOI}
\PrintBackRefs{\CurrentBib}

\bibitem [\protect \citeauthoryear {%
Suzuki%
\ \BBA {} Cavanagh%
}{%
Suzuki%
\ \BBA {} Cavanagh%
}{%
{\protect \APACyear {1998}}%
}]{%
suzuki1998shape}
\APACinsertmetastar {%
suzuki1998shape}%
\begin{APACrefauthors}%
Suzuki, S.%
\BCBT {}\ \BBA {} Cavanagh, P.%
\end{APACrefauthors}%
\unskip\
\newblock
\APACrefYearMonthDay{1998}{}{}.
\newblock
{\BBOQ}\APACrefatitle {A shape-contrast effect for briefly presented stimuli.}
  {A shape-contrast effect for briefly presented stimuli.}{\BBCQ}
\newblock
\APACjournalVolNumPages{Journal of Experimental Psychology: Human Perception
  and Performance}{24}{5}{1315}.
\newblock
\begin{APACrefDOI} \doi{https://doi.org/10.1037/0096-1523.24.5.1315}
  \end{APACrefDOI}
\PrintBackRefs{\CurrentBib}

\bibitem [\protect \citeauthoryear {%
Thornton%
\ \BBA {} Maurice%
}{%
Thornton%
\ \BBA {} Maurice%
}{%
{\protect \APACyear {1997}}%
}]{%
thornton1997physique}
\APACinsertmetastar {%
thornton1997physique}%
\begin{APACrefauthors}%
Thornton, B.%
\BCBT {}\ \BBA {} Maurice, J.%
\end{APACrefauthors}%
\unskip\
\newblock
\APACrefYearMonthDay{1997}{}{}.
\newblock
{\BBOQ}\APACrefatitle {Physique contrast effect: Adverse impact of idealized
  body images for women} {Physique contrast effect: Adverse impact of idealized
  body images for women}.{\BBCQ}
\newblock
\APACjournalVolNumPages{Sex Roles}{37}{5-6}{433--439}.
\newblock
\begin{APACrefDOI} \doi{https://doi.org/10.1023/A:1025609624848}
  \end{APACrefDOI}
\PrintBackRefs{\CurrentBib}

\bibitem [\protect \citeauthoryear {%
Tulving%
}{%
Tulving%
}{%
{\protect \APACyear {1981}}%
}]{%
tulving1981}
\APACinsertmetastar {%
tulving1981}%
\begin{APACrefauthors}%
Tulving, E.%
\end{APACrefauthors}%
\unskip\
\newblock
\APACrefYearMonthDay{1981}{}{}.
\newblock
{\BBOQ}\APACrefatitle {Similarity Relations in Recognition} {Similarity
  relations in recognition}.{\BBCQ}
\newblock
\APACjournalVolNumPages{Journal of Verbal Learning and Verbal
  Behavior}{20}{}{479-496}.
\PrintBackRefs{\CurrentBib}

\bibitem [\protect \citeauthoryear {%
Weiner%
}{%
Weiner%
}{%
{\protect \APACyear {1985}}%
}]{%
Weiner1985}
\APACinsertmetastar {%
Weiner1985}%
\begin{APACrefauthors}%
Weiner, B.%
\end{APACrefauthors}%
\unskip\
\newblock
\APACrefYearMonthDay{1985}{}{}.
\newblock
{\BBOQ}\APACrefatitle {{An Attributional Theory of Achievement Motivation and
  Emotion}} {{An Attributional Theory of Achievement Motivation and
  Emotion}}.{\BBCQ}
\newblock
\APACjournalVolNumPages{Psychological Review}{92}{4}{548--573}.
\newblock
\begin{APACrefDOI} \doi{https://doi.org/10.1037/0033-295X.92.4.548}
  \end{APACrefDOI}
\PrintBackRefs{\CurrentBib}

\bibitem [\protect \citeauthoryear {%
Wexley%
, Yukl%
, Kovacs%
\BCBL {}\ \BBA {} Sanders%
}{%
Wexley%
\ \protect \BOthers {.}}{%
{\protect \APACyear {1972}}%
}]{%
wexley1972importance}
\APACinsertmetastar {%
wexley1972importance}%
\begin{APACrefauthors}%
Wexley, K\BPBI N.%
, Yukl, G\BPBI A.%
, Kovacs, S\BPBI Z.%
\BCBL {}\ \BBA {} Sanders, R\BPBI E.%
\end{APACrefauthors}%
\unskip\
\newblock
\APACrefYearMonthDay{1972}{}{}.
\newblock
{\BBOQ}\APACrefatitle {Importance of contrast effects in employment
  interviews.} {Importance of contrast effects in employment
  interviews.}{\BBCQ}
\newblock
\APACjournalVolNumPages{Journal of Applied Psychology}{56}{1}{45-48}.
\newblock
\begin{APACrefDOI} \doi{https://doi.org/10.1037/h0032132} \end{APACrefDOI}
\PrintBackRefs{\CurrentBib}

\bibitem [\protect \citeauthoryear {%
Wickens%
\ \BBA {} Dixon%
}{%
Wickens%
\ \BBA {} Dixon%
}{%
{\protect \APACyear {2007}}%
}]{%
Wickens:2007hm}
\APACinsertmetastar {%
Wickens:2007hm}%
\begin{APACrefauthors}%
Wickens, C\BPBI D.%
\BCBT {}\ \BBA {} Dixon, S\BPBI R.%
\end{APACrefauthors}%
\unskip\
\newblock
\APACrefYearMonthDay{2007}{}{}.
\newblock
{\BBOQ}\APACrefatitle {The benefits of imperfect diagnostic automation: a
  synthesis of the literature} {The benefits of imperfect diagnostic
  automation: a synthesis of the literature}.{\BBCQ}
\newblock
\APACjournalVolNumPages{Theoretical Issues in Ergonomics
  Science}{8}{3}{201--212}.
\newblock
\begin{APACrefDOI} \doi{https://doi.org/10.1080/14639220500370105}
  \end{APACrefDOI}
\PrintBackRefs{\CurrentBib}

\bibitem [\protect \citeauthoryear {%
Wickens%
\ \protect \BOthers {.}}{%
Wickens%
\ \protect \BOthers {.}}{%
{\protect \APACyear {2009}}%
}]{%
Wickens:2009False}
\APACinsertmetastar {%
Wickens:2009False}%
\begin{APACrefauthors}%
Wickens, C\BPBI D.%
, Rice, S.%
, Keller, D.%
, Hutchins, S.%
, Hughes, J.%
\BCBL {}\ \BBA {} Clayton, K.%
\end{APACrefauthors}%
\unskip\
\newblock
\APACrefYearMonthDay{2009}{}{}.
\newblock
{\BBOQ}\APACrefatitle {False Alerts in Air Traffic Control Conflict Alerting
  System: Is There a "Cry Wolf" Effect?} {False alerts in air traffic control
  conflict alerting system: Is there a "cry wolf" effect?}{\BBCQ}
\newblock
\APACjournalVolNumPages{Human Factors}{51}{4}{446--462}.
\newblock
\begin{APACrefDOI} \doi{https://doi.org/10.1177/0018720809344720}
  \end{APACrefDOI}
\PrintBackRefs{\CurrentBib}

\bibitem [\protect \citeauthoryear {%
Wixted%
\ \BBA {} Wells%
}{%
Wixted%
\ \BBA {} Wells%
}{%
{\protect \APACyear {2017}}%
}]{%
Wixted2017}
\APACinsertmetastar {%
Wixted2017}%
\begin{APACrefauthors}%
Wixted, J\BPBI T.%
\BCBT {}\ \BBA {} Wells, G\BPBI L.%
\end{APACrefauthors}%
\unskip\
\newblock
\APACrefYearMonthDay{2017}{}{}.
\newblock
{\BBOQ}\APACrefatitle {{The Relationship Between Eyewitness Confidence and
  Identification Accuracy: A New Synthesis}} {{The Relationship Between
  Eyewitness Confidence and Identification Accuracy: A New Synthesis}}.{\BBCQ}
\newblock
\APACjournalVolNumPages{Psychological Science in the Public
  Interest}{18}{1}{10--65}.
\newblock
\begin{APACrefDOI} \doi{https://doi.org/10.1177/1529100616686966}
  \end{APACrefDOI}
\PrintBackRefs{\CurrentBib}

\bibitem [\protect \citeauthoryear {%
Wood%
}{%
Wood%
}{%
{\protect \APACyear {1978}}%
}]{%
wood1978knew}
\APACinsertmetastar {%
wood1978knew}%
\begin{APACrefauthors}%
Wood, G.%
\end{APACrefauthors}%
\unskip\
\newblock
\APACrefYearMonthDay{1978}{}{}.
\newblock
{\BBOQ}\APACrefatitle {The knew-it-all-along effect.} {The knew-it-all-along
  effect.}{\BBCQ}
\newblock
\APACjournalVolNumPages{Journal of Experimental Psychology: Human Perception
  and Performance}{4}{2}{345}.
\newblock
\begin{APACrefDOI} \doi{https://doi.org/10.1037/0096-1523.4.2.345}
  \end{APACrefDOI}
\PrintBackRefs{\CurrentBib}

\bibitem [\protect \citeauthoryear {%
Yang%
, Guo%
\BCBL {}\ \BBA {} Schemanske%
}{%
Yang%
\ \protect \BOthers {.}}{%
{\protect \APACyear {To appear}}%
}]{%
Yang2021}
\APACinsertmetastar {%
Yang2021}%
\begin{APACrefauthors}%
Yang, X\BPBI J.%
, Guo, Y.%
\BCBL {}\ \BBA {} Schemanske, C.%
\end{APACrefauthors}%
\unskip\
\newblock
\APACrefYearMonthDay{To appear}{}{}.
\newblock
{\BBOQ}\APACrefatitle {{From Trust to Trust Dynamics: Combining Empirical and
  Computational Approaches to Model and Predict Trust Dynamics in
  Human-Autonomy Interaction}} {{From Trust to Trust Dynamics: Combining
  Empirical and Computational Approaches to Model and Predict Trust Dynamics in
  Human-Autonomy Interaction}}.{\BBCQ}
\newblock
\BIn{} V\BPBI G.~Duffy, S\BPBI J.~Landry, J\BPBI D.~Lee\BCBL {}\ \BBA {} N\BPBI
  A.~Stanton\ (\BEDS), \APACrefbtitle {Human-Automation Interaction:
  Transportation.} {Human-automation interaction: Transportation.}
\PrintBackRefs{\CurrentBib}

\bibitem [\protect \citeauthoryear {%
Yang%
, Unhelkar%
, Li%
\BCBL {}\ \BBA {} Shah%
}{%
Yang%
\ \protect \BOthers {.}}{%
{\protect \APACyear {2017}}%
}]{%
Yang:2017:EEU:2909824.3020230}
\APACinsertmetastar {%
Yang:2017:EEU:2909824.3020230}%
\begin{APACrefauthors}%
Yang, X\BPBI J.%
, Unhelkar, V\BPBI V.%
, Li, K.%
\BCBL {}\ \BBA {} Shah, J\BPBI A.%
\end{APACrefauthors}%
\unskip\
\newblock
\APACrefYearMonthDay{2017}{}{}.
\newblock
{\BBOQ}\APACrefatitle {Evaluating Effects of User Experience and System
  Transparency on Trust in Automation} {Evaluating effects of user experience
  and system transparency on trust in automation}.{\BBCQ}
\newblock
\BIn{} \APACrefbtitle {{Proceedings of the 2017 ACM/IEEE International
  Conference on Human-Robot Interaction}} {{Proceedings of the 2017 ACM/IEEE
  International Conference on Human-Robot Interaction}}\ (\BPGS\ 408--416).
\newblock
\APACaddressPublisher{New York, NY, USA}{ACM}.
\newblock
\begin{APACrefDOI} \doi{https://doi.org/10.1145/2909824.3020230}
  \end{APACrefDOI}
\PrintBackRefs{\CurrentBib}

\bibitem [\protect \citeauthoryear {%
Yang%
, Wickens%
\BCBL {}\ \BBA {} H{\"{o}}ltt{\"{a}}-Otto%
}{%
Yang%
\ \protect \BOthers {.}}{%
{\protect \APACyear {2016}}%
}]{%
Yang:2016}
\APACinsertmetastar {%
Yang:2016}%
\begin{APACrefauthors}%
Yang, X\BPBI J.%
, Wickens, C\BPBI D.%
\BCBL {}\ \BBA {} H{\"{o}}ltt{\"{a}}-Otto, K.%
\end{APACrefauthors}%
\unskip\
\newblock
\APACrefYearMonthDay{2016}{}{}.
\newblock
{\BBOQ}\APACrefatitle {How users adjust trust in automation: Contrast effect
  and hindsight bias} {How users adjust trust in automation: Contrast effect
  and hindsight bias}.{\BBCQ}
\newblock
\BIn{} \APACrefbtitle {{Proceedings of the Human Factors and Ergonomics Society
  Annual Meeting}} {{Proceedings of the Human Factors and Ergonomics Society
  Annual Meeting}}\ (\BVOL~60, \BPGS\ 196--200).
\newblock
\begin{APACrefDOI} \doi{https://doi.org/10.1177/1541931213601044}
  \end{APACrefDOI}
\PrintBackRefs{\CurrentBib}

\bibitem [\protect \citeauthoryear {%
Zhang%
\ \BBA {} Yang%
}{%
Zhang%
\ \BBA {} Yang%
}{%
{\protect \APACyear {2017}}%
}]{%
Zhang2017:HFES}
\APACinsertmetastar {%
Zhang2017:HFES}%
\begin{APACrefauthors}%
Zhang, M\BPBI Y.%
\BCBT {}\ \BBA {} Yang, X\BPBI J.%
\end{APACrefauthors}%
\unskip\
\newblock
\APACrefYearMonthDay{2017}{}{}.
\newblock
{\BBOQ}\APACrefatitle {Evaluating effects of workload on trust in automation,
  attention allocation and dual-task performance} {Evaluating effects of
  workload on trust in automation, attention allocation and dual-task
  performance}.{\BBCQ}
\newblock
\BIn{} \APACrefbtitle {{Proceedings of the Human Factors and Ergonomics Society
  Annual Meeting}} {{Proceedings of the Human Factors and Ergonomics Society
  Annual Meeting}}\ (\BVOL~61, \BPGS\ 1799--1803).
\newblock
\begin{APACrefDOI} \doi{https://doi.org/10.1177/1541931213601932}
  \end{APACrefDOI}
\PrintBackRefs{\CurrentBib}

\end{thebibliography}

\newpage
\section{Biographies}
\textbf{X. Jessie Yang} is an Assistant Professor in the Department of Industrial and Operations Engineering at the University of Michigan Ann Arbor. She obtained a PhD in Mechanical and Aerospace Engineering (Human Factors) from Nanyang Technological University, Singapore in 2014.\\

\textbf{Christopher Schemanske} is an MSE student in the Department of Industrial and Operations Engineering at the University of Michigan Ann Arbor. When the present work was conducted, he was an undergraduate student in the same department.\\

\textbf{Christine Searle} is a MS student at the Robotics Institute, University of Michigan Ann Arbor. She obtained a BA in Computer Science and Psychology in 2014 from Indiana University Bloomington.\\







\end{document}